\documentclass[12pt,preprint]{aastex}

\usepackage{natbib}

\shorttitle{The EBAI Project}
\shortauthors{Pr\v sa, Guinan, Devinney, DeGeorge, Bradstreet, Giammarco, Alcock, Engle}

\newcommand {\phoebe}{{\tt PHOEBE }}
\newcommand {\phoebents}{{\tt PHOEBE}} 

\newcommand {\ebai}{{\tt EBAI }}
\newcommand {\ebaints}{{\tt EBAI}} 

\begin{document}

\title{Artificial Intelligence Approach to the Determination of Physical Properties of Eclipsing Binaries. I. The EBAI Project.}

\author{
	A.~Pr\v sa\altaffilmark{1,3},
	E.~F.~Guinan\altaffilmark{1},
	E.~J.~Devinney\altaffilmark{1},
	M.~DeGeorge\altaffilmark{1},
	D.~H.~Bradstreet\altaffilmark{2},
	J.~M.~Giammarco\altaffilmark{2},
	C.~R.~Alcock\altaffilmark{4},
	S.~G.~Engle\altaffilmark{1}
}
\altaffiltext{1}{Villanova University, Dept.~Astron.~Astrophys., 800 E Lancaster Ave, Villanova, PA 19085, USA}
\altaffiltext{2}{Eastern University, Dept.~of Physical Science, 1300 Eagle Rd, St.~Davids, PA 19087}
\altaffiltext{3}{University of Ljubljana, Dept.~of Physics, Jadranska 19, SI-1000 Ljubljana, Slovenia, EU}
\altaffiltext{4}{Harvard-Smithsonian Center for Astrophysics, 60 Garden St, Cambridge, MA 02138}

\begin{abstract}
Achieving maximum scientific results from the overwhelming volume of astronomical data to be acquired over the next few decades will demand novel, fully automatic methods of data analysis.  Artificial intelligence approaches hold great promise in contributing to this goal.  Here we apply neural network learning technology to the specific domain of eclipsing binary (EB) stars, of which only some hundreds have been rigorously analyzed, but whose numbers will reach millions in a decade. Well-analyzed EBs are a prime source of astrophysical information whose growth rate is at present limited by the need for human interaction with each EB data-set, principally in determining a starting solution for subsequent rigorous analysis.  We describe the artificial neural network (ANN) approach which is able to surmount this human bottleneck and permit EB-based astrophysical information to keep pace with future data rates. The ANN, following training on a sample of 33,235 model light curves, outputs a set of approximate model parameters ($T_2/T_1$, $(R_1+R_2)/a$, $e \sin \omega$, $e \cos \omega$, and $\sin i$) for each input light curve data-set. The whole sample is processed in just a few seconds on a single 2GHz CPU. The obtained parameters can then be readily passed to sophisticated modeling engines. We also describe a novel method \emph{polyfit} for pre-processing observational light curves before inputting their data to the ANN and present the results and analysis of testing the approach on synthetic data and on real data including fifty binaries from the Catalog and Atlas of Eclipsing Binaries (CALEB) database and 2580 light curves from OGLE survey data.  The success rate, defined by less than a 10\% error in the network output parameter values, is approximately 90\% for the OGLE sample and close to 100\% for the CALEB sample -- sufficient for a reliable statistical analysis. The code is made available to the public. Our approach is applicable to EB light curves of all classes; this first paper in the Eclipsing Binaries via Artificial Intelligence (\ebaints) series focuses on detached EBs, which is the class most challenging for this approach.
\end{abstract}

\keywords{methods: data analysis --- methods: numerical --- binaries: eclipsing --- stars: fundamental parameters}

\section{Introduction}

Over the past decade advances in observational technologies, computers and eclipsing binary (EB) analysis codes (e.g., WD code: \citet{wd1971,wd1993,wd2007}) have enabled the accumulation of an impressive sample of fundamental stellar data.   Careful analysis of EB light curves has produced fundamental stellar properties, tests of stellar evolution theories, accurate distances within the Galaxy and to external galaxies, as well as providing tests of stellar structure models and general relativity \citep[see e.g.][]{guinan2007}.  Despite the importance of these astrophysical results, only some hundreds of EBs have been subject to the requisite analysis and the cumulative results populate the astrophysical parameter space sparsely. Current standard practice requires significant human interaction with EB light curve data, particularly in the initial stages of analysis, which defines a "rate-determining step" in the process generating these astrophysical results.

By 2020, the observational bounty from ground- and space-based programs such as Optical Gravitational Lensing Experiment \citep[OGLE;][]{udalski1997}, Exp\'erience de Recherche d'Objets Sombres \citep[EROS;][]{pd1998}, All Sky Automated Survey \citep[ASAS;][]{pojmanski2002}, Pan-Starrs \citep{kaiser2002}, Large Synoptic Survey Telescope \citep[LSST;][]{tyson2002}, and their space counterparts Hipparcos \citep{perryman1997}, Kepler \citep{borucki2004} and Gaia \citep{perryman2001} will include millions of new EB light curves, even catching some EBs in fleeting stages of stellar evolution. Powerful and mature codes for light curve analysis stand ready to mine this enormous and rich vein of new astrophysical information, and pioneering steps in automating these tools have already been taken by \citet{devor2005,tamuz2006,mazeh2006} (see \citet{prsa2007} for a concise overview of current automated techniques). A key issue is to efficiently determine an initial set of model parameters for every light curve as input to the automated analysis process. Starting values have been conventionally supplied by the analyst/astronomer using expert knowledge typically guided by checks with light curve synthesis codes, a time-intensive process that is certainly out of the question for the coming fire-hose of EB data.

Approaches to automating light curve solutions have taken various forms to date. \citet{ww2001,ww2002}, in their work to establish the best distance indicators among detached and semi-detached binaries in the Small Magellanic Cloud, obtained starting parameters for the rigorous WD model by comparing each candidate light curve with a set of template model light curves, sending the best match to an automated version of the WD differential corrector program DC. This could be computationally prohibitive to apply to the expected large future data-sets, even if clever pruning reduced the number of comparison templates.  Employing less rigorous physical models, of course, is one approach to computational efficiency. \citet{devor2005} provides a critical discussion of an automated pipeline employing a simple model of spherical stars without tidal or reflection physics, whose starting values are obtained from an initial guess and then refined using a downhill simplex method with simulated annealing. \citet{tamuz2006} employ the EBOP ellipsoidal model \citep{popper1981}.  Using this engine, they arrive at initial solutions after a combination of grid search, gradient descent and geometrical analysis of the LC.

The challenge is therefore to gain the advantages of a sophisticated model yet keep processing time limited. The Eclipsing Binaries via Artificial Intelligence (\ebaints) project introduces artificial neural networks (ANNs) that are trained on the rigorous WD physical model and are computationally extremely efficient, towards fully automating the solution process for EBs.

In this first paper in the \ebai series we describe the basic ANN concepts and procedures for applying ANNs to detached EBs.  We present results of applying the trained ANN to a set of 10,000 synthetic detached EBs, to 50 detached real world binaries from the Catalog and AtLas of Eclipsing Binaries (CALEB\footnote{CALEB is maintained by D.~Bradstreet at Eastern University -- see {\tt http://caleb.eastern.edu}.}), and to the set of 2580 OGLE LMC binaries \citep{wyrzykowski2003} classified as detached. Subsequent papers will deal similarly with semi-detached systems and overcontact EBs, and address automated light curve classification.

\section{The EBAI project: concept and implementation}

\begin{figure}
\includegraphics[width=16cm]{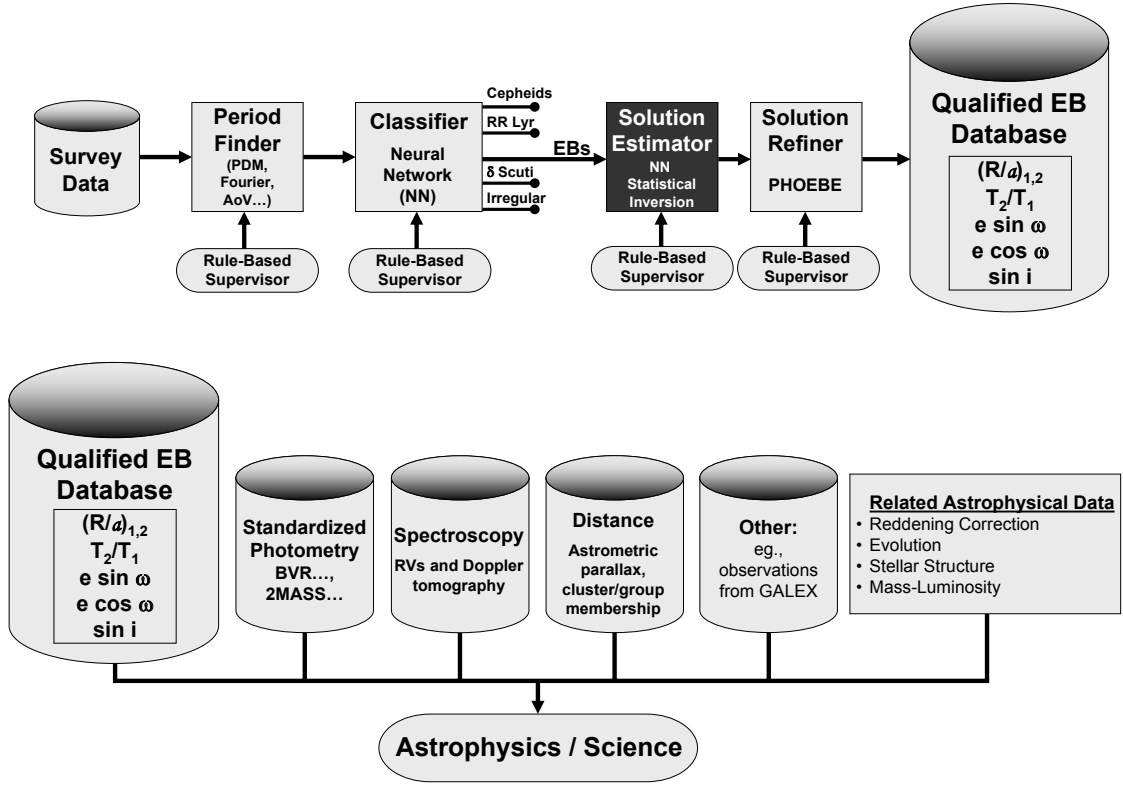} \\
\caption{Intelligent Data Pipeline (IDP). Complete survey data is piped through a period finder algorithm that is controlled by a rule-based system. All variable sources are then passed to the ANN-based classifier. Light curves consistent with EB signatures are passed to the Solution Estimator block (focus of this paper). The output from the Solution Estimator is input to the sophisticated WD-based engine \phoebents. For detached EBs the IDP produces a catalog of 5 principal parameters -- $T_2/T_1$, $\rho_1+\rho_2$, $e \sin \omega$, $e \cos \omega$, and $\sin i$. These are complemented by other survey/mission data and follow-up observations to obtain full-scale astrophysical results.} \label{idp}
\end{figure}

Advances in Artificial Intelligence (AI) and the continued operation of Moore's law that predicts doubling of the processing power every two years have created the opportunity for significant progress in solving the types of problems that are limited by the lack of human capital. A new approach, the \emph{Intelligent Data Pipeline} (IDP), is being prototyped in the domain of EBs which uses AI techniques to operate autonomously on large observational data-sets to produce results of astrophysical value. The IDP is designed to handle the complete process of variable discovery, classification of variability and management of the solution process for the discovered EBs (\citealt{devinney2005,devinney2006}; cf.~Fig.~\ref{idp}). The IDP employs ANNs in the processing modules, while the supervisory knowledge, now implicit in humans, is encoded in control modules as rules appropriate for each processing module. The supervisory modules have the task of keeping the process on track and providing physically meaningful results through each phase of the processing pipeline. This paper concentrates on the \emph{Solution Estimator} block: given the set of classified input light curves, it finds a set of physical and geometric parameters that best match the input data.

\subsection{Artificial Neural Networks}

\begin{figure}
\begin{center}
\includegraphics[width=12cm]{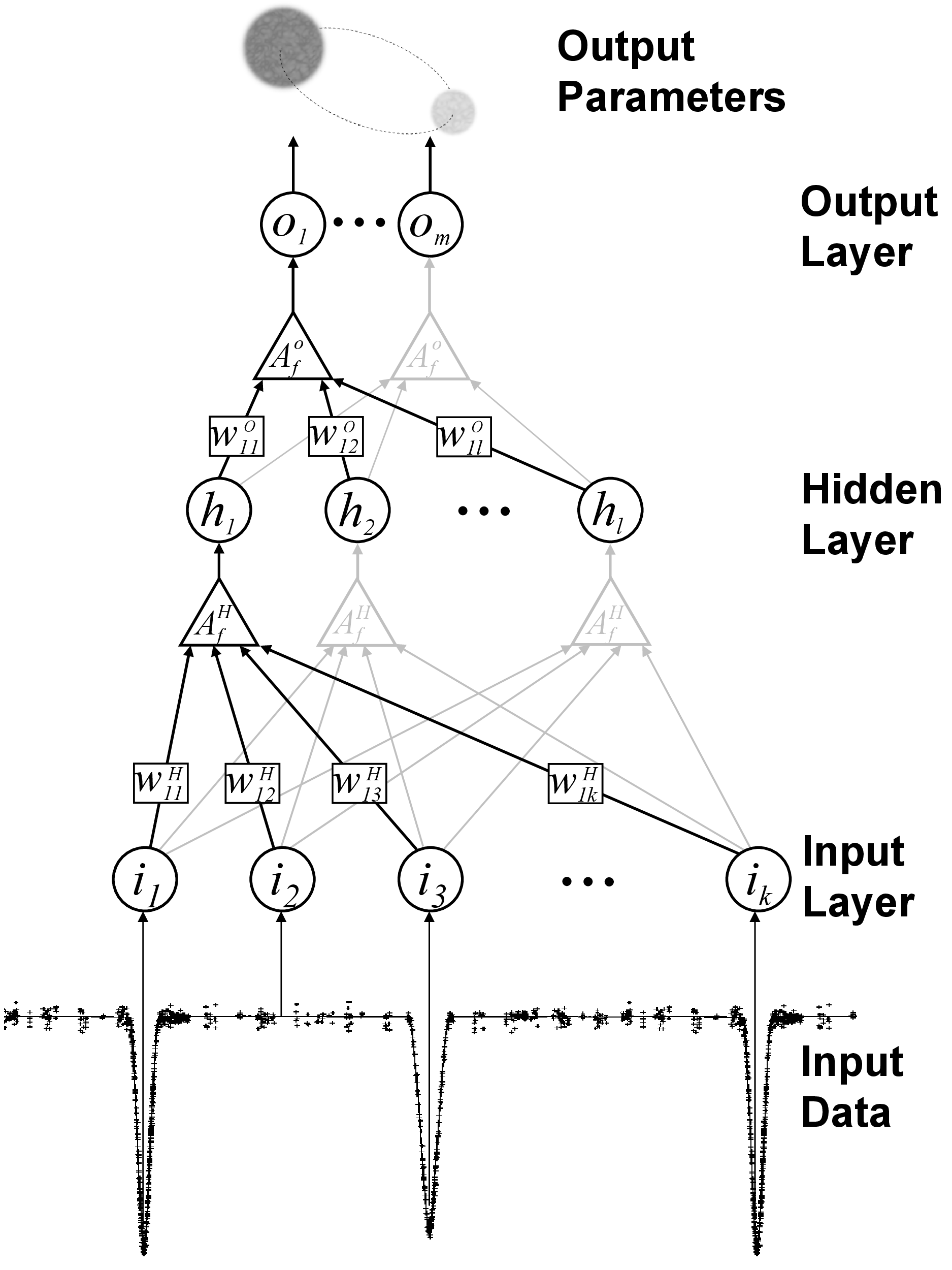} \\
\end{center}
\caption{The topology of a three-layer Artificial Neural Network (ANN). Processing units (nodes) on the input layer are populated with flux observations that are equidistant in phase. A weighted sum of values on the input layer, $y_j = \sum_k w_{jk}^H i_k$, stimulates each unit, $j = 1 \dots l$, on the hidden layer. The amount of stimulus is determined by the activation function $A_f^H$. In our case this is a sigmoid function, $A_f^H (y_j) = 1/[1+\exp(-(y_j-\mu)/\tau)]$, with parameters $\mu$ and $\tau$ chosen in such a way that $A_f^H$ is mapped onto the $[-1,1]$ interval. Processing units on the hidden layer are then populated with $h_j = A_f^H (y_j)$. An analogous propagation takes place between the hidden layer and the output layer, using the same type of the activation function. The output layer of a trained network then contains the network's best guess at physical parameters of the data that have populated the input layer. The ANN is thus a \emph{mapping} from observations to the physical parameters of the observed system.} \label{ann}
\end{figure}

An artificial neural network is a simple construct: it is a stack of interconnected \emph{layers} (cf.~Fig.~\ref{ann}). Each layer is an array of processing elements called \emph{units}. These units propagate the signal between layers by \emph{weighted connections}. The units perform non-linear \emph{mapping} of input data to output parameters.

The input to the network is a light curve -- an array of photometric data points at equidistant phase intervals. These data are the stimulus of the \emph{input} layer: each unit on the layer acquires a value of the given input array element. Once the input layer is populated, propagation to the \emph{hidden} layer begins. The stimulus of the given unit on the hidden layer is a weighted sum of outputs from units on the current layer. This stimulus is then passed through a non-linear \emph{activation function} that determines the extent of stimulation of the given unit. There is a connection from every unit on the input layer to every unit on the hidden layer, and each of these connections has a corresponding weight assigned to it. Once the signal has been propagated to the hidden layer, the propagation continues to the \emph{output} layer: each unit acquires a value that is a weighted sum of outputs from units on the hidden layer, passed through the activation function. The network thus \emph{maps} its input to its output by propagating the stimulus via weighted connections that are passed through activation functions. Given the layer-to-layer connection weights, the propagation is basically a matter of summation and multiplication. The power comes from the ability of networks to provide \emph{non-linear} mapping between their input and their output by using non-linear activation functions. The most commonly used activation functions are sigmoid curves, i.e.~functions of the type $f(x) = 1/[1+\exp(-(x-\mu)/\tau)]$.

The goal of ANNs is to have output layer values that have some representative significance of the input array. In our case the output is an array of model parameters that correspond to the input light curve. Since the output values depend on the connection weights, the task is to determine these weights. This is where back-propagation comes into play. Assume we have a sample of several thousand \emph{exemplars} (input arrays for which we know the corresponding physical parameters) -- obtained, for example, by computing theoretical light curves, or by using real data with reliable model solutions. For each exemplar we perform forward-propagation and compare the results output by the network with the true ones. We then \emph{modify} the weights so that the discrepancy between the results is reduced for the whole sample. This iterative procedure is called the \emph{training} or \emph{machine learning} phase. It is the only computation-intensive block and it is performed only once. Once the weights are determined and the network reliably reproduces the expected results, the network is ready to \emph{recognize} input never seen before. Hundreds of thousands of light curves can subsequently be processed in a matter of seconds on a single CPU.

Depending on their construction, ANNs can serve either as classifiers or regression calculators. The classification mode features \emph{bins} that are representative of a group of objects instead of continuous output parameters. The regression mode, on the other hand, imposes no limits on the domain of parameter values and provides a non-linear mapping between the input layer and the output layer. The IDP classifier is an example of the classifying ANN, whereas the solution estimator ANN (the main topic of this paper) is a typical regression ANN.

The network described here, a basic three-layer back-propagating network (BPN), is remarkably robust for a diversity of non-linear problems. Different network topologies, such as multiple hidden layers, as well as more complicated connection strategies, advanced training approaches and other variations in general do not bring significant improvements to the basic model. For a thorough discussion on neural networks please refer to authoritative books such as \citet{freeman1991}.

\subsection{Choice of ANN principal parameters}

Solving the inverse problem for EBs (determining the set of model parameters that best fit the observations) is notoriously difficult because parameter inter-correlations and degeneracy are inherent to the problem and cannot be avoided. Thus considerable care must be taken when selecting the network's output parameters. For example, if we focus on \emph{detached} EB light curves, parameters such as the mass ratio and semi-major axis cannot be obtained. Consequently, absolute-scale parameters such as the radii and individual surface potentials cannot be obtained either. Moreover, if there is only one input light curve, there is no information on the absolute values of individual components' temperatures.

The role of ANNs is to provide \emph{estimates} for parameter values as a starting point for specialized minimization methods, such as NMS or Powell's method \citep{prsa2005, prsa2006}, therefore all second order parameters such as gravity darkening, reflection effect, or spots, can also be ignored. In fact, because of degeneracy it seems that, out of 40 or so parameters involved in computing a light curve, there is virtually no parameter that would be ideal to appear on the ANN's output list. To stay true to the idea of artificial intelligence, and to overcome these difficulties, we must take a step back: what guides humans to recognize (and implicitly classify and analyze) a light curve? ANN is a \emph{recognition} algorithm, so it has to connect to that which triggers our recognition capability.

\begin{figure}
\begin{center}
\includegraphics[width=14cm]{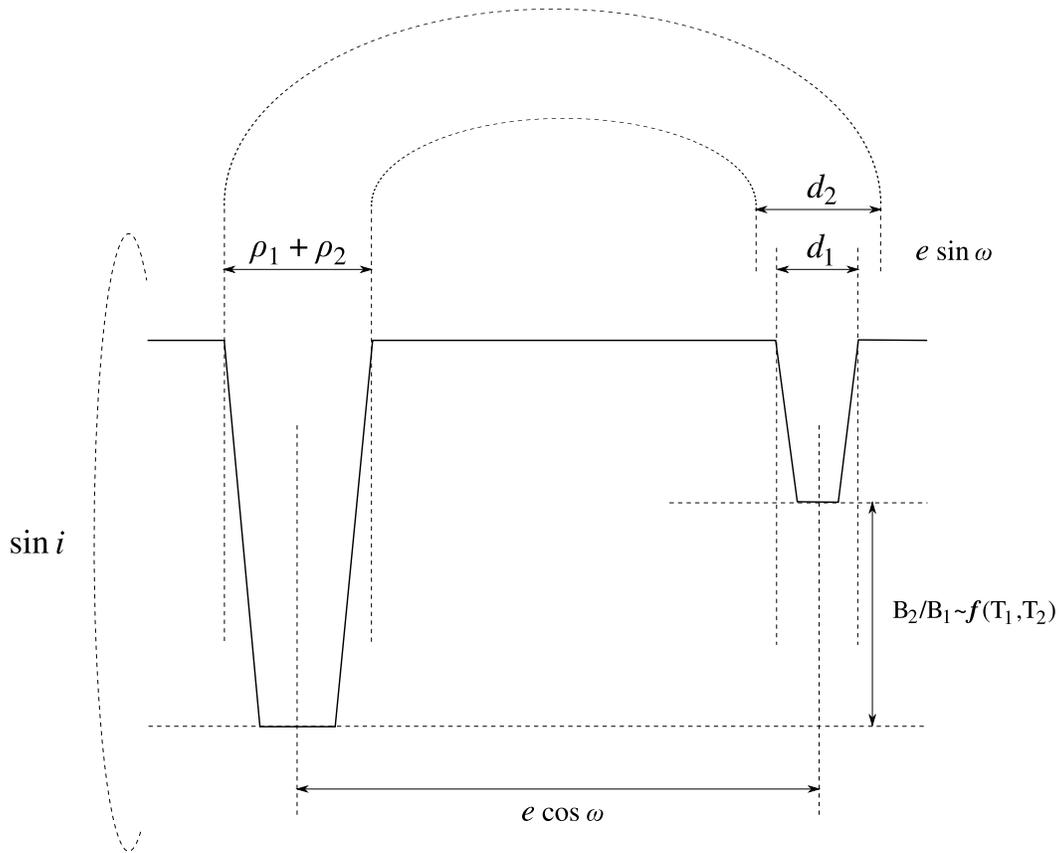} \\
\end{center}
\caption{Principal parameters of a light curve that bridge our subjective recognition with impersonal parameters of the EB system.} \label{principal}
\end{figure}

What guides humans to recognize light curves are the ratios and the geometry. Accordingly, we sought a set of parameters that would act as a bridge between the features that trigger our subjective recognition, and physical and geometrical parameters of the system. Fig.~\ref{principal} schematically presents the shape of a light curve and the parameters that determine such a shape. Namely, the surface brightness ratio $B_2/B_1$ determines the ratio of depths of both eclipses; the sum of relative radii $\rho_1+\rho_2$ determines eclipse width; eccentricity $e$ and the argument of periastron $\omega$ determine the separation between the eclipses and also the ratio of their widths; finally, the inclination $i$ determines the overall shape of the eclipse -- {\sf U} vs.~{\sf V}-shaped minima -- and the overall amplitude of the light curve. We have selected these indicators to form the five principal parameters of the ANN. They are summarized in Table \ref{principal_table}.

\begin{table}
\caption{Ranges of principal parameters for detached eclipsing binaries.} \label{principal_table}
\begin{center}
\begin{tabular}{lcl}
\hline \hline
Parameter:                 & Range:      & Description: \\
\hline
$\alpha = T_2/T_1$         & [\phantom{-}0.1, 1.0] & effective temperature ratio \\
$\beta = \rho_1 + \rho_2$  & [0.01, 0.5] & sum of relative radii, $\rho = R/a$ \\
$\gamma = e \sin \omega$   & [-0.3, 0.3] & radial eccentricity \\
$\delta = e \cos \omega$   & [-0.3, 0.3] & tangential eccentricity \\
$\epsilon = \sin i$        & [0.85, 1.0] & sine of the inclination \\
\hline
\end{tabular}
\end{center}
\end{table}

\begin{description}
\item[Effective temperature ratio.] Although the ratio of depths of both eclipses depends directly on the surface brightness ratio and is thus a complicated function of temperature, we approximate it here with the temperature ratio. The rationale behind this choice is threefold: 1) the equivalence between the temperature ratio and the surface brightness ratio increases as the deformation of stellar surfaces decreases and this approximation is thus valid for well detached binaries; 2) there exists a direct mapping from local temperature distribution over the stellar surface to surface brightness, but there is no direct mapping from surface brightness to local temperatures. To obtain the temperatures (that are input parameters to the model) from surface brightness ratio would thus imply a computationally intensive iterative scheme that would significantly prolong the overall run-time; and 3) since we are basing our analysis on a single light curve, there is no inherent way to calibrate the temperatures and thus no way to obtain the effective temperature of the binary. As consequence, even if we used surface brightness as the network's output parameter, we would still have to \emph{assume} one of the temperatures and therefore still suffer from systematic effects.
\item[Sum of relative radii.] The absolute scale of the system cannot be unambiguously obtained from a single light curve, nor can we typically get individual fractional radii. The reason for the latter is the degeneracy between the radii: any combination can reproduce the curve as long as their sum is kept constant. This degeneracy is broken in special cases, i.e.~total eclipses, pronounced proximity effects, or eccentric orbits, since the signature of individual radii is predominantly contained in the ingress and egress shape of the eclipse. For close detached, semi-detached, and overcontact systems this limitation does not exist. In general, however, the only quantity attainable from the width of the eclipses of well detached EBs is the sum of relative (fractional) radii $\rho_1+\rho_2 = (R_1+R_2)/a$.
\item[Radial component of eccentricity.] Were the eccentricity $e$ and argument of periastron $\omega$ to enter the training separately, problems would arise at small eccentricities; e.g., as $e$ gets smaller, $\omega$ becomes indeterminate. Furthermore, the orthogonalized components $e \sin \omega$ and $e \cos \omega$ have distinct effects on the shape of the light curve. Radial eccentricity $e \sin \omega$ is directly proportional to the ratio $(d_2-d_1)/(d_2+d_1)$ of the durations of both minima.
\item[Tangential component of eccentricity.] The tangential orthogonalized component $e \cos \omega$ is directly related to the displacement of the minima. This component is a first order effect, whereas the eclipse width change is a second order effect.
\item[Inclination.] Instead of using the inclination itself, it proves better to use $\sin i$ because of its geometric significance, that is the angle analogous to the sum of fractional radii that determines the shape and duration of the eclipse.
\end{description}

For the purpose of ANN, these parameters are linearly \emph{remapped} to the $[0.1,0.9]$ interval, values that are suitable for the activation function.

\subsection{The training data-set}

The goal of the \ebai project is to rapidly process large amounts of survey data. If the network is to perform successfully, the training sample has to be representative of those data. Although survey results are impressive because of the sheer number of the observed targets, they lack in accuracy, data diversity (i.e. simultaneous photometry, spectroscopy, astrometry) and often have poor phase coverage. We typically have a single photometric light curve in instrumental magnitudes with errors as great as $\sim 5\%$. There may be fewer than $\sim 100$ data points, and because of the typical vagaries of survey/mission operation (scanning laws) these do not cover the phase range uniformly. Well detached EBs might thus have only a few points in the eclipses, with practically flat quadratures. It is important to be aware of these limitations in the observed data, since we cannot expect ANNs to do more than a human could do interactively. From a poor light curve there remains a very limited amount of information that can be extracted reliably.

\begin{figure}
\includegraphics[width=15cm]{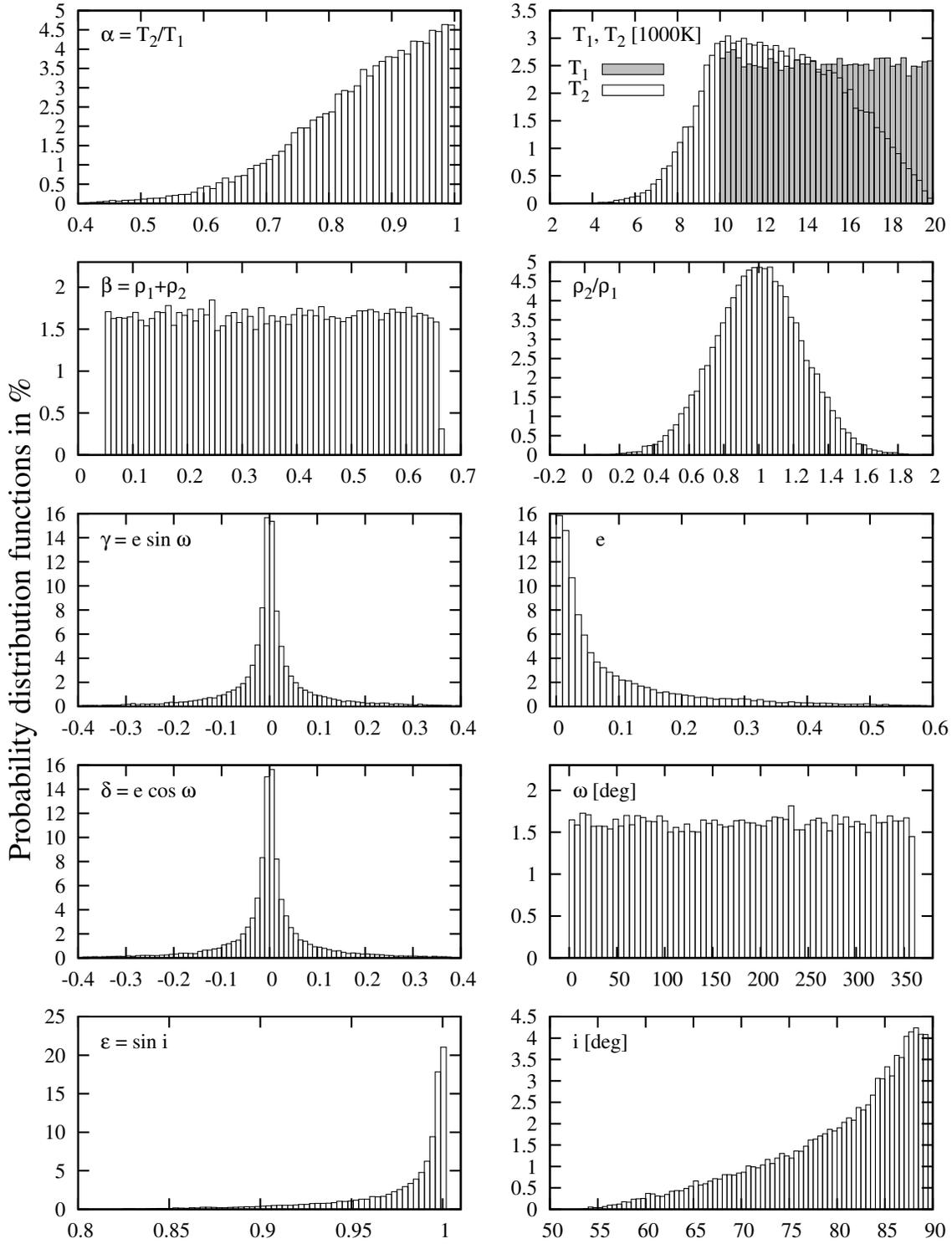} \\
\caption{Distributions of the principal parameters ($\alpha$ through $\epsilon$; left panels) and of the corresponding physical parameters ($T_1$ through $i$; right panels). A Monte-Carlo generator was used to create a training set of 33,235 synthetic light curves with parameters sampled randomly, according to these distribution functions.} \label{dists}
\end{figure}

Our training data-set is a representative version of the expected realistic data described above. We created an initial sample of 33,235 normalized light curves in the Cousins I band, each consisting of 200 data points distributed equidistantly in phase, and with added synthetic white noise. We used a Monte Carlo-based script in \phoebe \citep{phoebe} that randomly selected the values of parameters of detached eclipsing binaries according to the distribution functions depicted in Fig.~\ref{dists}. For eccentric binaries a phase shift $\Delta \Phi = (M_c + \omega - 1/4)/2\pi$, where $M_c$ is the mean anomaly at conjunction, was introduced so that the primary eclipse coincides with phase 0. The number of created light curves (33,235) was chosen so that it is close to $8^5 = 32768$ and a multiple of 23 to optimize use of a 24-node Beowulf cluster. The first criterion means that the 5-parameter grid resolution is $\sim 8\times$ the interval width: the determination of the argument of periastron, for example, has a resolution of $2\pi/8 = \pi/4$. Beyond that, the accuracy is limited by the ANN interpolation error. Of all parameters, the limiting interpolation capability of the network affects only the argument of periastron since its signature is present in only a small subset of the whole sample (EBs with significant eccentricity). Tests have shown that increasing the number of exemplars improves the determination of the argument of periastron significantly. Other parameters do not depend as strongly on the number of exemplars.

Although ANNs depend only weakly on the distribution functions of exemplar parameters, some care should be taken to avoid systematic effects due to the inherent degeneracy of EB light curves. For example, the initial training sample was created uniformly across all parameters and, although ANN performance was reasonable, it exhibited clear systematics in cases where light curves are not sensitive to parameter variation. This was most noticeable for eccentric binaries with arguments of periastron close to $\pm \pi/2$: eccentric orbits with minima separated by half the orbital period statistically overwhelmed circular orbits, causing the network to bias eccentric binaries with semi-major axes aligned with the line of sight to the circular binaries. To reduce this and similar effects, the distribution functions are selected so that systematics due to degeneracy are minimized. Thus, $\alpha$ follows a gaussian distribution function $\mathcal N(1.0, 0.2)$; $\beta$ is sampled uniformly between 0.01 and 0.5, with the ratio of fractional radii distributed by $\mathcal N(1.0,1.3)$; $\gamma$ and $\delta$ follow an exponential distribution $\mathcal E(0,0.05)$ in $e$ and a uniform distribution in $\omega$; finally, $\varepsilon$ follows a uniform distribution in $i$ with the built-in cut-off point $i_\mathrm{min} = i_\mathrm{grazing} - 1^\circ$, where $i_\mathrm{grazing}$ is the critical inclination for eclipses to occur. Limb darkening values are interpolated for each generated light curve from \citet{vanhamme1993} tables, and gravity darkening coefficient $\beta$ is set to 1.0 for radiative envelopes ($T_\mathrm{eff} > 7500$\,K; \citealt{vonzeipel1924}) and 0.32 for convective envelopes ($T_\mathrm{eff} < 7500$\,K; \citealt{lucy1967}). A canonical value of the mass ratio $q=1.0$ is used for all curves. In the absence of other types of data (spectra or RVs), the effects of the degeneracy can never be completely removed, only reduced.

\subsection{Fitting a smooth curve to observed data} \label{polyfit_section}

\begin{figure}
\includegraphics[angle=-90,width=16cm]{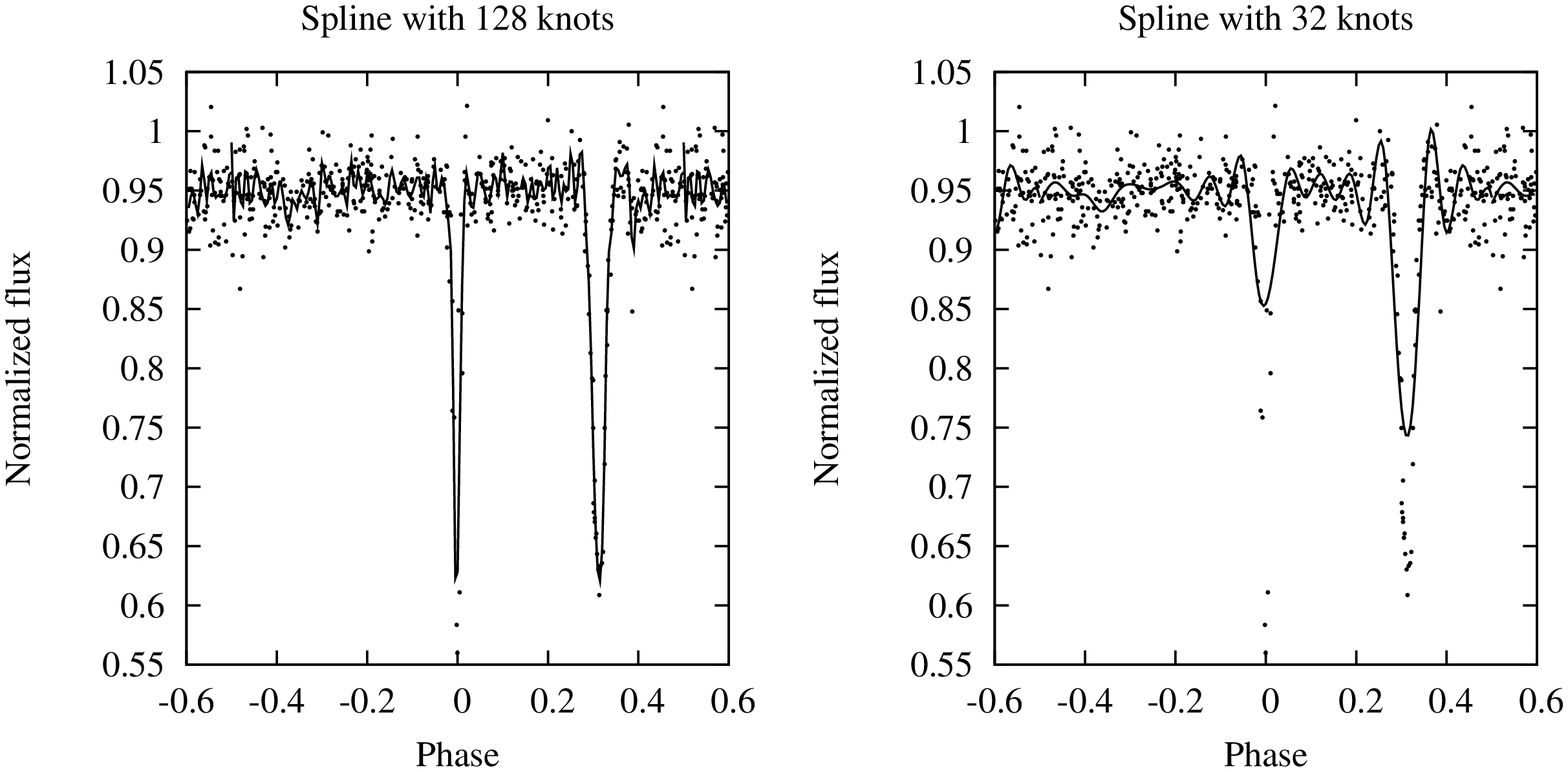} \\
\caption{Examples of a cubic B-spline fit to OGLE data. In the first case (128 knot spline) the overall shape of the light curve is reproduced, but at the expense of large oscillations. In the second case (32 knot spline) the fit to eclipses is deteriorated, and the oscillations still remain. Splines proved not well suited for fitting detached EB light curves.} \label{spline}
\end{figure}

The network expects its input to be given at 200 equidistant phases ranging from $-0.5$ to $0.5$, with phase 0 corresponding to the primary (deeper) minimum. To achieve this, the observed data points must be represented by a smooth analytic function that can later be sampled in equidistant phase points. In the special circumstances -- complete/uniform phase coverage and low noise, the data can simply be interpolated or binned into normals. Unfortunately, neither condition is usually satisfied by survey data. The task of fitting a smooth curve to such observed data proved quite challenging: common fitting functions like Fourier series, orthogonal polynomials (Legendre, Chebyshev, etc), or splines do a poor job on real-world detached binary light curves. Due to the function regularity requirement (functions must be connected and differentiable in every point), data noise induces strong oscillations in the fitting functions, making them virtually useless (cf.~Fig.~\ref{spline}). We therefore devised a dedicated algorithm for fitting a smooth curve to detached binary data. Our algorithm \emph{polyfit} fits a polynomial chain $\mathcal P(x)$ of piecewise smooth $n$-th order polynomials connected at a given set of \emph{knots}. The two key ideas of the algorithm are 1) to abandon the requirement of differentiability of $\mathcal P(x)$ at knots and thus allow the polynomial chain to be broken, and 2) to vary the position of the knots iteratively and thus relax the system to the nearest minimum. The formal description of the algorithm with the implementation details are given in Appendix \ref{polyfit_appendix}. Fig.~\ref{polyfit_examples} shows some examples of the algorithm's performance on OGLE data-set.

\begin{figure}
\begin{center}
\includegraphics[width=14cm]{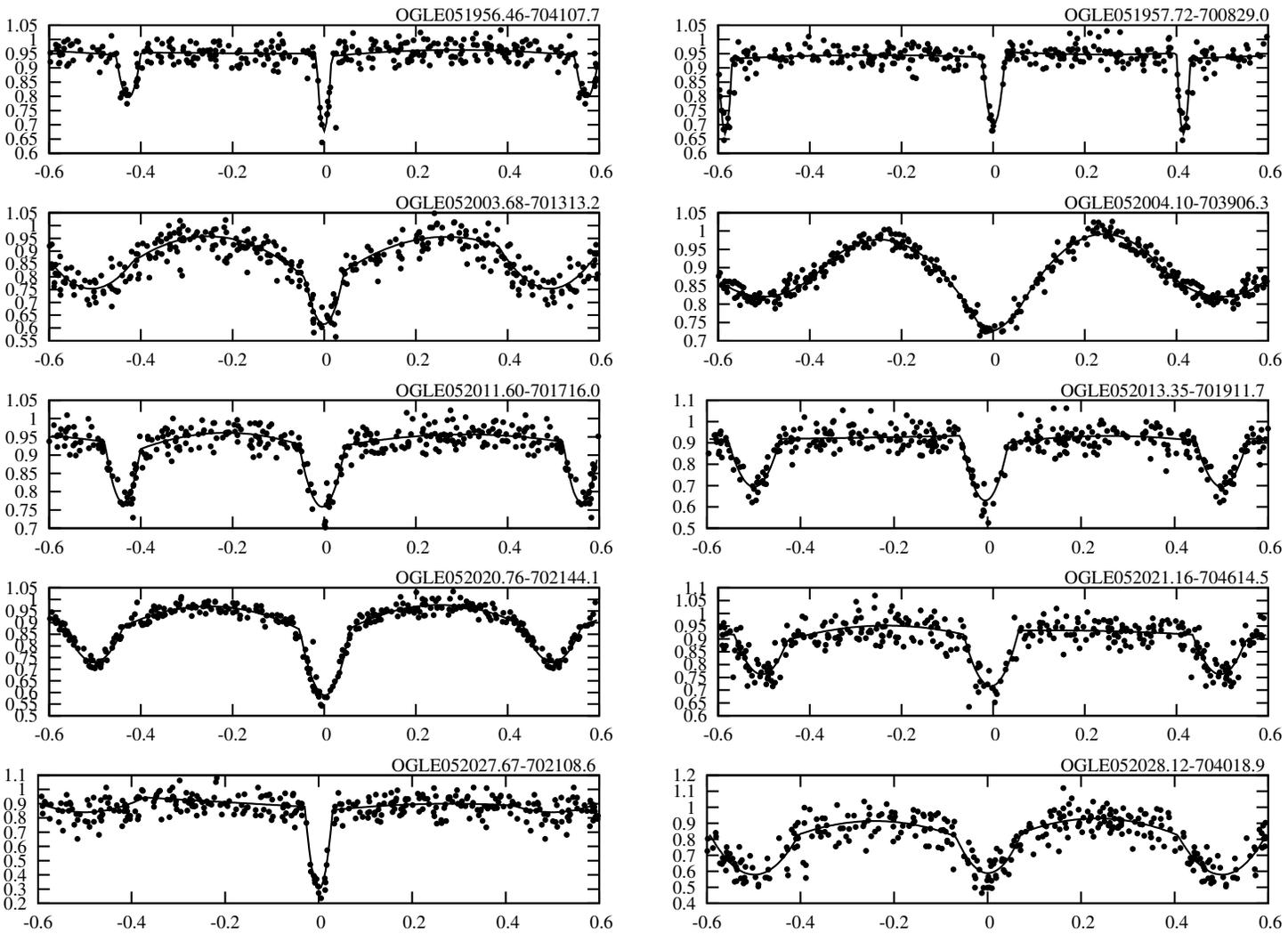} \\
\end{center}
\caption{Example of \emph{polyfit} algorithm fitting applied to 10 OGLE LMC SC21 light curves (filled circles) with the corresponding \emph{polyfit} solutions (solid lines). The complete catalog of \emph{polyfit} solutions may be retrieved electronically from {\tt http://www.eclipsingbinaries.org}.} \label{polyfit_examples}
\end{figure}

\subsection{ANN sensitivity to light levels}

The generated exemplar light curve fluxes are normalized to the sum of both components' passband luminosities per steradian's worth of area, namely $f_\mathrm{norm} = 1 \equiv (L_1+L_2)/4\pi$, which assigns unity flux to \emph{spherical} stars of the given luminosity. Thus, the quarter-phase flux will be unity for well detached binaries and close to unity for deformed stars because of ellipsoidal variations and reflection effect. The rationale behind this choice of normalization lies in the nature of observed light curves: were we to normalize synthetic light curves to maximum light, we would have to do the same for observed light curves, a poor choice because we would be normalizing to outliers in the observed data.

\begin{figure}
\begin{center}
\includegraphics[height=14cm,angle=-90]{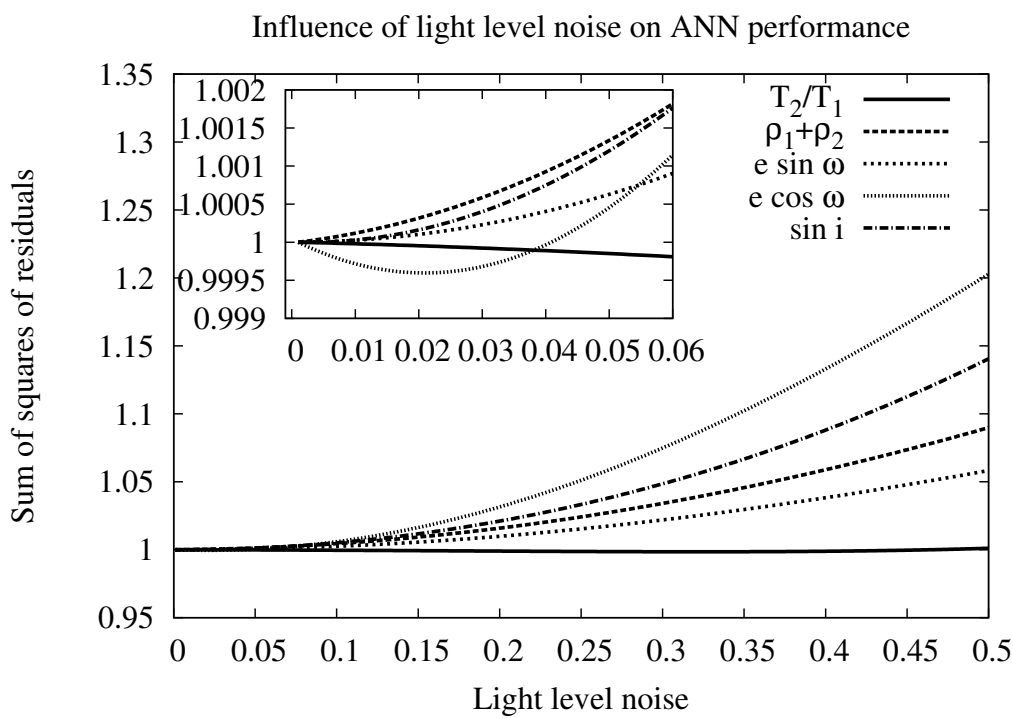} \\
\end{center}
\caption{Neural network performance on randomly normalized data. The depicted range of $\sigma$ = 0--50\% (0--6\% at the inset) in random displacement clearly shows that the network performance is independent of the input normalization level: 5\% error in normalization propagates to less than 0.2\% in the sum of squares of residuals, which in turn implies even smaller errors in terms of parameter accuracies, negligible for all practical purposes. The initial decreasing trend in $T_2/T_1$ and $e \cos \omega$ is a consequence of the 0.5\% jitter added to the exemplars during the learning phase.} \label{levnoise}
\end{figure}

The question is, then, how to normalize observed data so that light levels are consistent with those of the exemplars. To do that properly, we would need to know the amount of distortion of each component, an unknown which the network is supposed to yield. It turns out that this apparent problem is not a problem at all: neural networks predominantly learn the \emph{shape} of the light curves rather than normalization. Fig.~\ref{levnoise} depicts the dependence of the sum of squares of residuals on light level for 5000 exemplars. Varying each exemplar's normalization level according to the given $\sigma$ (ranging from 0\% to 50\%), the data were propagated through the network and the change in the sum of squares of residuals was observed. Since we normalize all light curves to 1.0 at quarter-phase, this introduces $\sim$5\% error in normalization for significantly distorted binaries. This error corresponds to less than 0.2\% in the sum of squares, which in turn corresponds to less than 0.01\% in parameter errors, negligible for all practical purposes.

\subsection{Selection of the optimal ANN topology}

\begin{figure}
\begin{center}
\includegraphics[height=14cm,angle=-90]{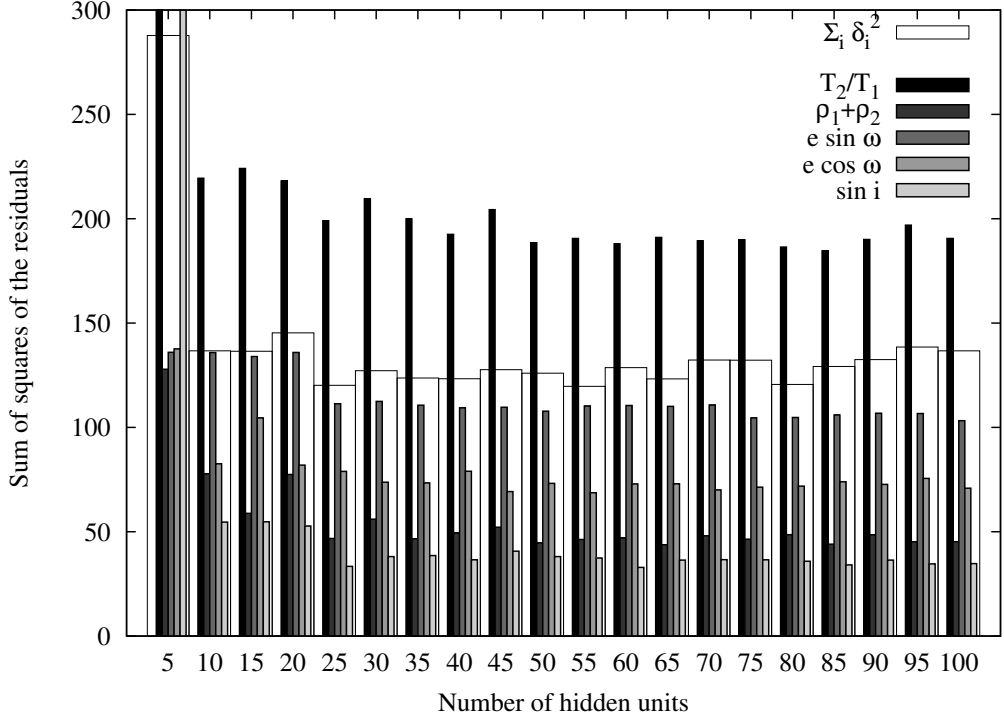} \\
\end{center}
\caption{Network performance as a function of the number of hidden units. Filled bars correspond to the sum of squares of the residuals for individual parameters ($\alpha$: green; $\beta$: blue; $\gamma$: purple; $\delta$: cyan; $\epsilon$: yellow) of 33235 exemplars. The outlined red bar depicts a rescaled sum of squares of the residuals for all 5 parameters, i.e.~the actual cost function of the network. The network performs reasonably well for 25 to 65 hidden units; given that the training time cost grows quadratically with the number of hidden units, the smallest still satisfactory number should be adopted. In accordance with the above diagram and following the detailed analysis of the residual trends (deviations from the normal noise distribution), we set the number of hidden units to 40.} \label{hidden_units}
\end{figure}

Three-layer back-propagation neural networks are well known for their robustness and applicability on a wide range of computational problems \citep{freeman1991}. EBs proved to be no exception: tests with two or more hidden layers had shown that there is no obvious benefit for convergence stability. The real challenge was to determine the optimal sizing of the network topology. The input layer is constrained by the input data (200 data points at equidistant phases), and the output layer is constrained by the output parameters. Thus, the topological parameter to be explored is the number of hidden units. We trained the network identically for different numbers of hidden units (Fig.~\ref{hidden_units}), which yielded the cost function value for individual parameters (solid bars) and the complete output layer (outline bar) as a function of the number of hidden units. With few hidden units, there is insufficient freedom for the network to learn to map $e \sin \omega$ and $e \cos \omega$, whereas too many hidden units cause the network to oscillate by learning noise traits. The optimal choice corresponds to the number of hidden units where the cost-function dependency levels off, at around 40 units. This value was selected for the final network topology.

\begin{figure}
\begin{center}
\includegraphics[height=14cm,angle=-90]{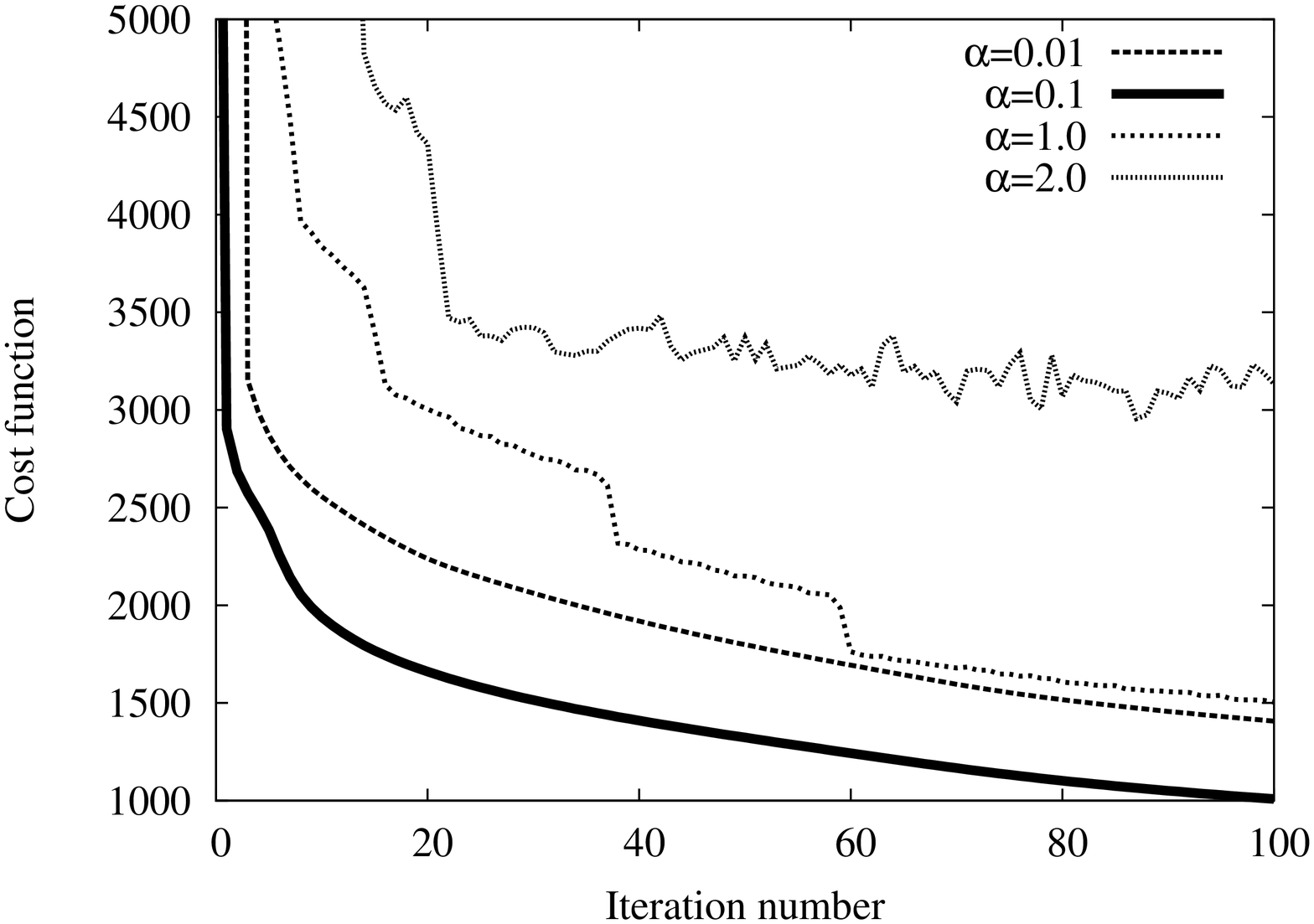} \\
\end{center}
\caption{Comparison between learning curves for different learning rate parameters ($\alpha=0.01$, 0.1, 1.0, and 2.0). For small values of $\alpha$ the learning curve is smooth, but requires additional iterations to converge. For large values of $\alpha$ the learning curve oscillates as parameter values overshoot the targeted minimum. After a comprehensive search over a fine-grained range of learning rate parameters (only 4 are depicted here for clarity) we find that the optimal value is $\alpha=0.1$.} \label{lrp}
\end{figure}

The network learning rate parameter $\alpha$ determines the step size along the direction of the negative gradient during back-propagation. If $\alpha$ is too small, learning requires an increased number of iterations. If it is too large, the correction overshoots the oscillates about the target value. Fig.~\ref{lrp} depicts learning curves for $\alpha$ ranging from 0.01 to 2.0. The value $\alpha = 0.1$ proved best for the adopted network topology, and it was held fixed\footnote{There are back-propagation training algorithms that increase the learning rate parameter with the number of iterations, but we found no obvious necessity for implementing such a strategy.} during learning.

\section{ANN performance and results}

The principal benefit of ANNs is speed; a trained network typically takes seconds on an average computer to process even the largest EB databases. However, the most important question is the reliability of the parameters reproduced by the network. After training the network with 33,235 exemplars, it was tested on a distinct synthetic sample of 10,000 detached EB light curves. The second test was performed on the CALEB database of real-world binaries, where we extracted 50 detached EBs and passed them through the network. For these we could compare the known values of model parameters from the sample with parameters yielded by the network. After making sure the ANN performance was satisfactory, we submitted 2580 detached EBs from the OGLE database for which we had no prior solutions. We present the results of the analysis below.

\subsection{Network training} \label{nettraining}

All training light curves were pre-processed by the \emph{polyfit} algorithm. Although the equidistant phase requirement can be easily achieved for synthetic data, neural networks perform poorly if the training set is fundamentally different from the unknown data-set despite their apparent similarity. This requirement was met by fitting a polynomial chain to model light curves and training the network on such a pre-processed sample. Another merit of pre-processing the sample is a reduction of systematic error: if a polynomial chain fit exhibits systematic effects on an observed light curve, it will likely exhibit the same systematic effects on the model light curve.

\begin{figure}
\begin{center}
\includegraphics[height=14cm,angle=-90]{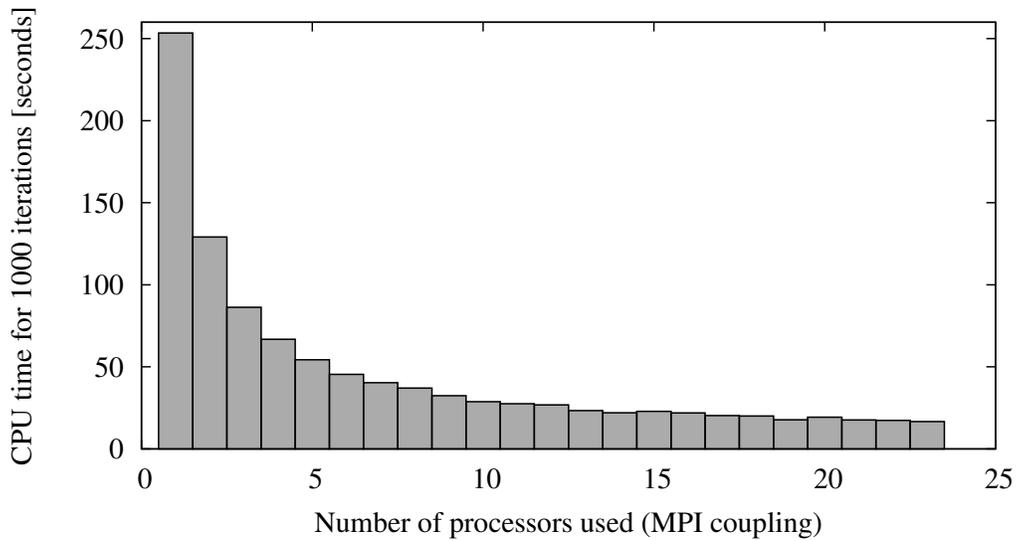} \\
\end{center}
\caption{Back-propagation speed-up resulting from multi-processor parallelization. The framework used is Message Passing Interface (MPI) as implemented by the OpenMPI library ({\tt http://www.open-mpi.org}). The algorithm was implemented on a dedicated 24-node 2GHz Opteron cluster. The net yield is $\sim 15\times$ speed-up with respect to the serial uni-processor implementation.} \label{parallelization}
\end{figure}

The back-propagation stage (ANN training) is the most computationally intensive block of the algorithm. It is inherently serial, since the weight matrices are updated one exemplar at a time. Given the network topology and a large number of exemplars, it would typically take a very long time (days, even weeks) to calibrate and train the network adequately. For example, the time required to train a network on a single 2GHz processor (250\,s for 1000 iterations on a single CPU, thus $\sim$1.45 days for 500,000 iterations; cf.~Figs.~\ref{parallelization}, \ref{learning_curve}) equals the time required to compute $\sim$12,500 eccentric light curves ($\sim$10\,s per eccentric light curve computation). Yet this comparison does not take into account the overwhelming computational time to calibrate network topology (in our case we trained at least 50 different networks), which in most circumstances dominates the overall processor time. To overcome this, we devised a parallel version of the back-propagation algorithm that partitions the sample among available processors in a Beowulf cluster. Each processor performs back-propagation on a chunk of the sample, and communicates the locally modified weight matrices back to the main unit. The main unit then assembles the final matrices by weighted averaging and broadcasts them to compute nodes for the next iteration. If the chunks are sufficiently large, there will be only a slight penalty for not performing a rigorous gradient computation. Moreover, there is an unexpected benefit: the learning process is better able to avoid local minima because of the perturbation of the steepest descent. Fig.~\ref{parallelization} shows the net speed-up as a function of the number of processors: running on 24 nodes, the speed gain is $\sim 15\times$, reducing 2 weeks worth of processor time to less than 1 day.

\begin{figure}
\begin{center}
\includegraphics[height=14cm,angle=-90]{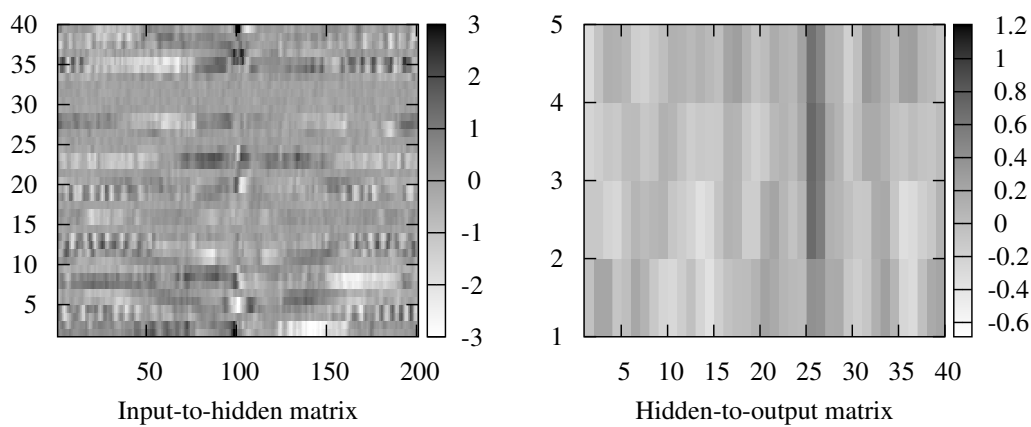} \\
\end{center}
\caption{Weight matrices. The I$\to$H matrix (left) plays a dominant role in mapping general LC features. There are pronounced peaks around the 100-th unit that corresponds to phase 0, and around the 0-th and 200-th units that correspond to phase 0.5. It is interesting to observe the transverse patterns in the I$\to$H matrix, indicating that the network not only learns the phase position of the LC features, but also the variation of slope along the phase. The H$\to$O matrix (right), on the other hand, serves for fine-tuning of the mapping. It does not exhibit any obvious form and the weights change throughout convergence both in magnitude and in distribution.} \label{weight_matrices}
\end{figure}

\begin{figure}
\begin{center}
\includegraphics[height=\textwidth,angle=-90]{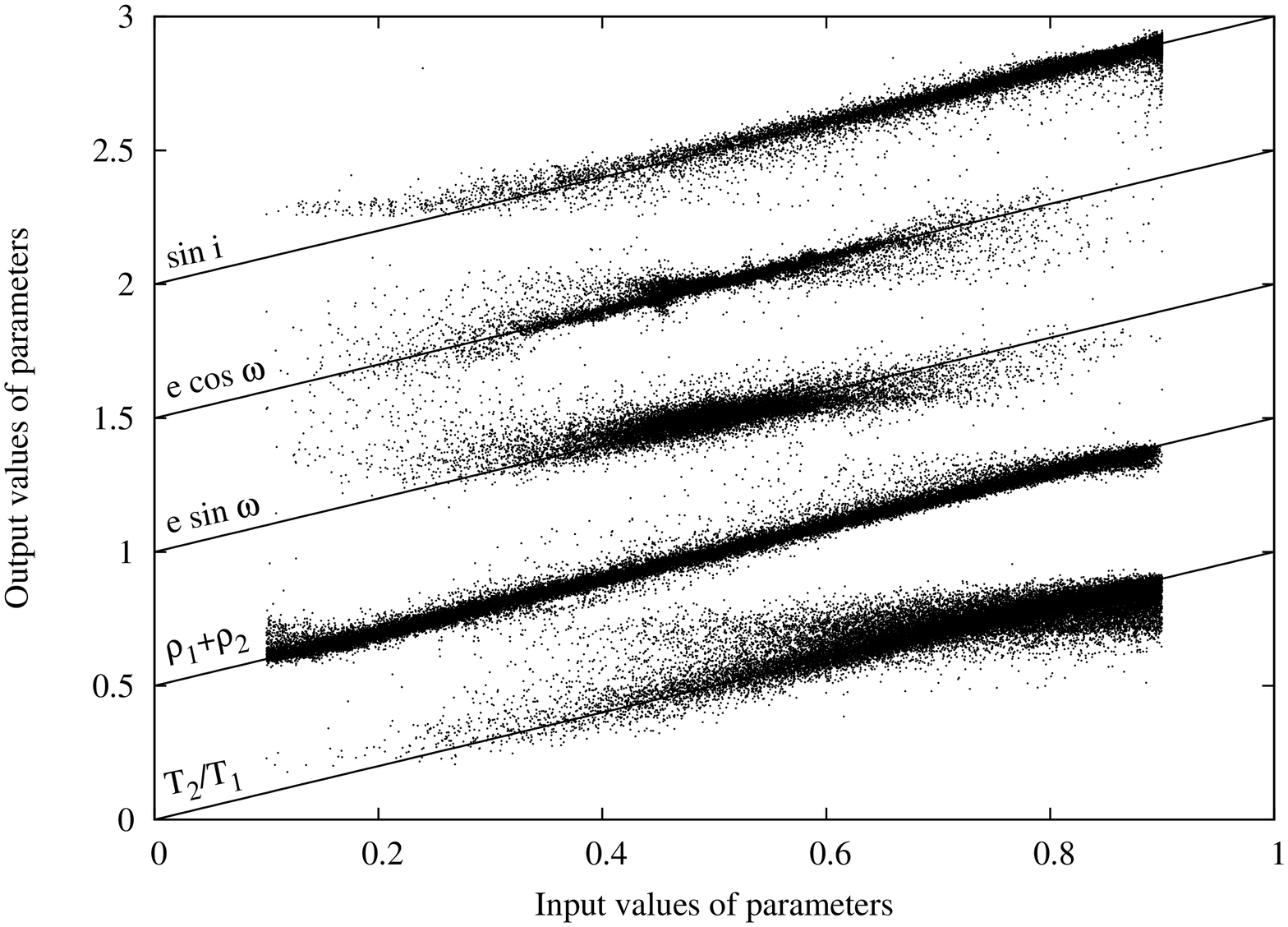} \\
\end{center}
\caption{The results of neural network training on 33,235 exemplars -- light curves generated by the WD engine. The correlation between input values and recognized output values is shown for each parameter and offset by 0.5 for clarity, with parameter values being linearly mapped to the $[0.1,0.9]$ interval for activation function efficiency. Labeled guidelines representing ideal performance are provided for easier comparison. Strong systematics are evident in the $T_2/T_1$ parameter, due to the poor approximation of the surface brightness ratio by $T_2/T_1$. Fortunately, temperature ratio is the parameter with most rapid DC convergence (demonstrated in Section \ref{caleb_main}) and a single DC iteration is usually necessary to correct for this problem. The remaining trends and areas of increased scatter are either statistically insignificant ($<5\%$ in $90\%$ of all cases; cf.~Fig.~\ref{results_diffs}), or are caused by a well-understood degeneracy that is discussed at length in the text.} \label{results}
\end{figure}

The network with 200 input nodes, 40 hidden nodes, and 5 output nodes was trained for 500,000 iterations with a learning rate parameter $\alpha = 0.1$. Variable white noise with $\sigma$ ranging from 0.005 to 0.02 was added to the sample of 33,235 synthetic light curves. Exemplars were created from these light curves by \emph{polyfit} pre-processing. Fig.~\ref{learning_curve} shows the learning curve: sum of squares of the residuals on the output layer as function of iteration number. The curve is monotonically decreasing, with the convergence rate falling below $10^{-5}$/iteration at around the 470000-th iteration. We tested convergence stability for the total of 10 million iterations and noted no deterioration of the fit (i.e.~from over-training) nor significant improvement of the adopted result. Fig.~\ref{weight_matrices} depicts the transformation matrices from input to hidden (I$\to$H) layer (left), and from hidden to output (H$\to$O) layer (right). The I$\to$H matrix determines general mapping properties and projects eclipse patterns clearly; it does not change substantially after the first 10000 iterations. The H$\to$O matrix serves for fine-tuning; it continues changing as iterations proceed and shows no clear patterns that would relate to either layer in any obvious way. The results of training are depicted on Fig.~\ref{results}. The following traits for each parameter may be deduced from this figure:

\begin{figure}
\begin{center}
\includegraphics[height=14cm,angle=-90]{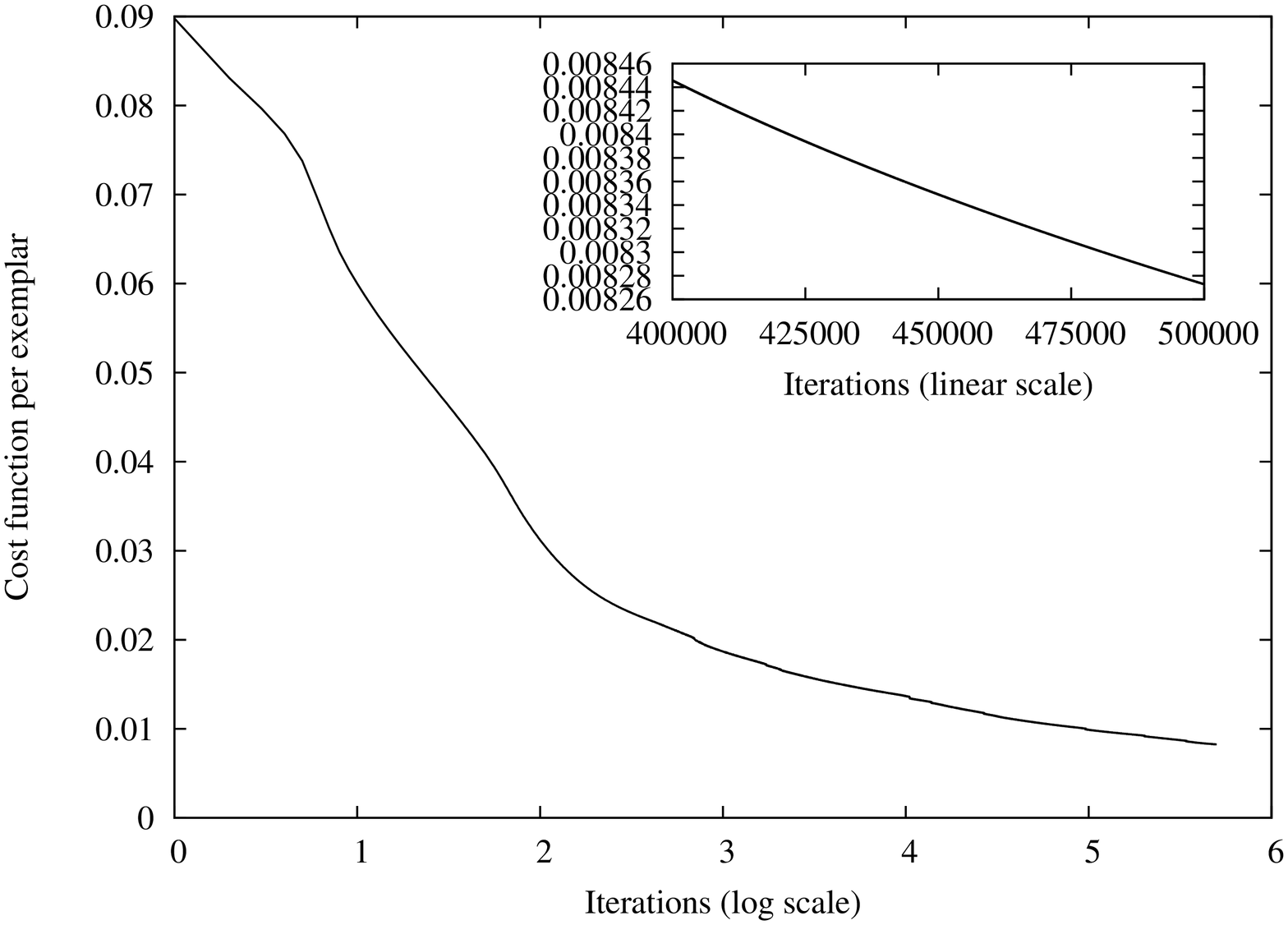} \\
\end{center}
\caption{Learning curve. The number of iterations on the main plot is depicted in log scale. Cost function is normalized per exemplar, i.e.~$\frac 1N \sum_i \sum_{p=1}^5 (o_p-c_p)^2$, where $N$ is the number of exemplars, $o_p$ and $c_p$ are input and output values of parameters, respectively, index $i$ goes over all exemplars and index $p$ goes over all output parameters. The inset depicts the final 100,000 iterations on a linear scale.} \label{learning_curve}
\end{figure}

\begin{figure}
\begin{center}
\includegraphics[height=\textwidth,angle=-90]{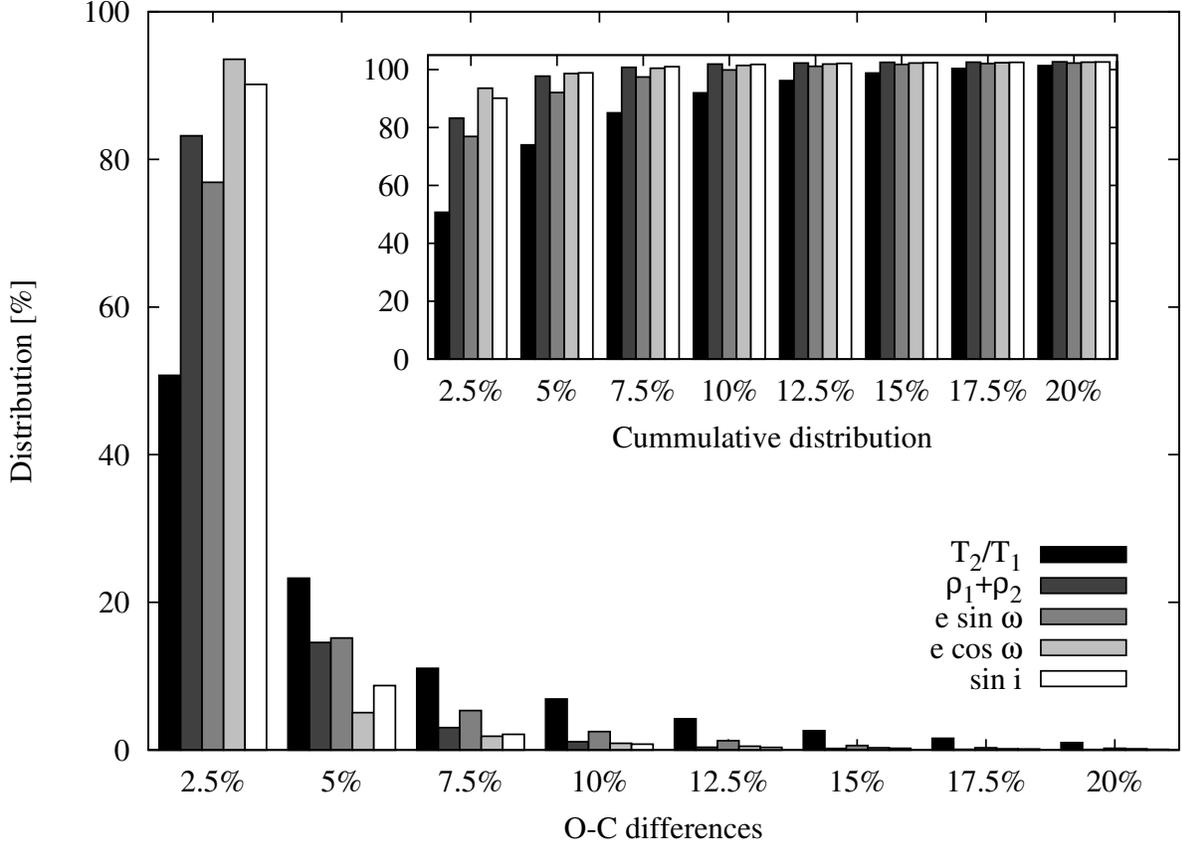} \\
\end{center}
\caption{Error distribution of ANN output parameters for the training sample. The bars depict the fraction of all EBs with errors between 0\% and 2.5\% (first bin), 2.5\% and 5\% (second bin), etc, for individual parameters. The $e \cos \omega$ parameter has the highest success recognition rate, and $T_2/T_1$ has the lowest success rate because of its approximate $B_2/B_1$ behavior. The inset shows the cumulative distribution for the same data: 90\% of all cases have errors in \emph{all} parameters smaller than 10\%.} \label{results_diffs}
\end{figure}

\begin{description}
\item [$T_2/T_1$.] Since surface brightness ratio, which is directly related to the ratio of depths of both eclipses, only approximately relates to the temperature ratio, it is no surprise that this parameter was the toughest to be reproduced by the network. Most systematic error occurs near $T_2/T_1 \approx 1$, where the model would expect equal depth eclipses. These of course depend also on the ratio of radii and on the radial component of eccentricity, so high scatter at the upper end of the distribution is not surprising. However, temperature ratio is the parameter with the most rapid DC convergence and it often takes only a single DC iteration to significantly improve it. For this parameter 49\% of the sample was reproduced to 2.5\% accuracy (cf.~Fig.~\ref{results_diffs}).
\item [$\rho_1+\rho_2$.] Neural network convergence was most rapid for the sum of fractional radii, cca.~1000 iterations required for ANN training. This arises because the eclipse width for detached binaries is a well behaved parameter, i.e.~the deformation of stellar surfaces does not have a significant influence on the duration of the eclipse. The fall-off at the lower end can be attributed to phase coverage: given the number of equidistant phase points, for wide EBs there are only a few pixels in the minima and these do not suffice for reliable discrimination. The upper end deviation, on the other hand, is due to degeneracy between the radii (surface potentials) and the inclination: the effect of slightly enlarging the radii or by slightly increasing the inclination on eclipse width is indistinguishable. For this parameter 81\% of the sample was reproduced to a 2.5\% accuracy.
\item [$e \sin \omega$.] The geometric significance of the radial component of eccentricity is in the ratio of the durations of both eclipses and is thus a second order effect. Although the network does learn the trend for weakly eccentric binaries, it systematically fails for strongly eccentric binaries. There are two reasons for this: 1) very eccentric orbits imply widely detached EBs typically with poor eclipse phase coverage, and 2) the number of cases with $e>0.2$ in our training sample is only $\sim 5$\%. We thoroughly tested the latter issue by training the network with 5000, 10,000, 20,000, and 33,235 exemplars; the improvement in the determination of $e \sin \omega$ was observed as anticipated. The phase coverage issue could be solved by increasing the number of phase points, but given the scatter in survey data-sets we anticipate the loss of accuracy on this account to be minimal. The bunching at mid-range is expected, as this corresponds to near-spherical orbits, and the majority of systems are in this range. For this parameter 75\% of the sample was reproduced to a 2.5\% accuracy.
\item [$e \cos \omega$.] Contrary to the radial component, the tangential component of eccentricity is better determined by the network. Its geometric significance lies in the phase displacement of both minima and is thus a first order effect. This parameter does not suffer from any obvious systematic effects. For this parameter 91\% of the sample was reproduced to a 2.5\% accuracy.
\item [$\sin i$.] The inclination proved to be recovered reliably for all practical purposes. The lower end discrepancy is of no practical importance because it corresponds to grazing eclipses and non-eclipsing binaries that constitute $\sim$1\% of the sample. The increased scatter on the upper end is due to the degeneracy between the radii and inclination already discussed. For this parameter 88\% of the sample was reproduced to a 2.5\% accuracy.
\end{description}

\subsection{Test 1: Synthetic sample of 10,000 EBs}

\begin{figure}
\begin{center}
\includegraphics[height=\textwidth,angle=-90]{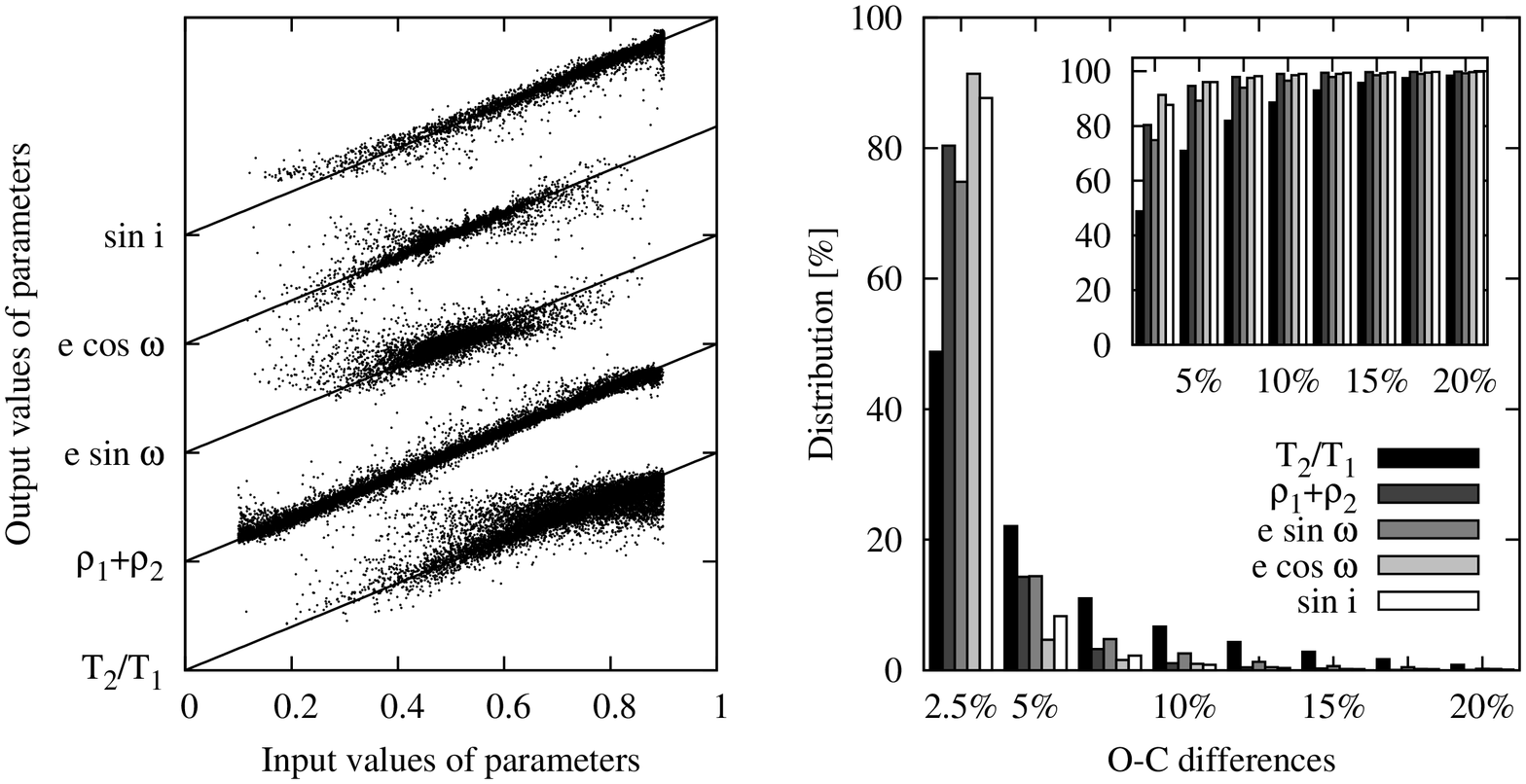} \\
\end{center}
\caption{Neural network recognition performance on 10,000 unknown synthetic LCs. The test sample was created using the same method as the training sample. Left: comparison between input and output values of individual parameters. Comparing against Fig.~\ref{results} shows that not only is the success rate of recognition comparable to that of the training set, but it also has the same underlying features identified and interpreted in the training phase. Right: distribution of differences (main graph) and their cumulative distribution (inset). Again, the same distribution as that of the training sample is seen.} \label{syntest}
\end{figure}

As a first test we evaluated the ANN's performance on a sample of 10,000 detached EB light curves distinct from the training set. The test sample was created using the same method as the training sample, following the same probability distribution functions for parameter values (cf.~Fig.~\ref{dists}) and adding variable amounts of white noise randomly chosen between $\sigma = 0.005$ and $\sigma = 0.02$. The sample was then pre-processed with \emph{polyfit} to obtain polynomial chain fits to the data. Fig.~\ref{syntest} shows the results of this test. The success rate of recognition is comparable to that of the learning sample (71\% of $T_2/T_1$, 95\% of $\rho_1+\rho_2$, 89\% of $e \sin \omega$, 96\% of $e \cos \omega$, and 96\% of $\sin i$ are determined to better than 5\%), and the underlying distribution of errors remains the same. This demonstrates the capability of the ANN to successfully recognize data it has never encountered before. In the worst case scenario (see cumulative inset in Fig.~\ref{syntest}) the network yields parameters accurate to 10\% for 90\% of the sample. This implies that the ANN output on unknown data is viable for statistical analysis and as input to sophisticated modeling engines for fine-tuning.

\subsection{Test 2: 50 real-world stars from the CALEB database} \label{caleb_main}

Further testing of our approach to automated light curve solution -- using a trained ANN followed by optional differential corrections -- on real-world detached binaries from the CALEB database. Manual inspection of 59 candidate light curves eliminated 9 as either semi-detached or overcontact, leaving 50 test binaries. Of these, 7 stars had published model parameters that did not exactly reproduce the observed LC, which is not surprising given the heterogeneous nature of CALEB data. For each binary, the chi-squared value of the model light curve generated from the ANN output parameters determined whether 0, 1, 2, or 3 passes would be made through the differential corrections program DC \citep{wd1971}. Table \ref{caleb_table} shows the results. For 22 systems, the ANN output generated sufficiently good fits that DC was not invoked. 1 DC iteration was performed for $\chi^2$ values between 1.5 and 2.0 (16 systems), 2 DC iterations for $\chi^2$ values between 2.0 and 2.5 (5 systems), and 3 DC iterations for $\chi^2$ values of 2.5 or higher (7 systems). Each of the remaining systems for which DC was invoked converged. Interestingly, for 7 of these systems the published parameters were discrepant with the results here, which suggests an additional benefit to a uniformly-based automated analysis.

\thispagestyle{empty}
\begin{deluxetable}{lccccccccccccccc} \label{caleb_table}
\tabletypesize{\tiny}
\rotate
\tablewidth{\textheight}
\tablecaption{Comparison between CALEB values of parameters ($\alpha_C$ through $\epsilon_C$), ANN values of parameters ($\alpha_N$ through $\epsilon_N$), and DC-improved values of parameters where applicable ($\alpha_{DC}$ through $\epsilon_{DC}$).}

\tablehead{
  \colhead{Star name:} &
  \colhead{$\alpha_C$}   & \colhead{$\alpha_N$}   & \colhead{$\alpha_{DC}$}   &
  \colhead{$\beta_C$}    & \colhead{$\beta_N$}    & \colhead{$\beta_{DC}$}    &
  \colhead{$\gamma_C$}   & \colhead{$\gamma_N$}   & \colhead{$\gamma_{DC}$}   &
  \colhead{$\delta_C$}   & \colhead{$\delta_N$}   & \colhead{$\delta_{DC}$}   &
  \colhead{$\epsilon_C$} & \colhead{$\epsilon_N$} & \colhead{$\epsilon_{DC}$}
}

\startdata
ADBoo   & 0.929030 & 0.863416 & n/a      & 0.298062 & 0.296898 & n/a      & 0.000000 & 0.001101 & n/a      & 0.000000 & 0.036864 & n/a      & 0.999263 & 0.996247 & n/a      \\
NNCep   & 0.972000 & 0.979644 & 0.872912 & 0.480238 & 0.471822 & 0.551348 & 0.000000 & 0.001251 & 0.001251 & 0.000000 & 0.015310 & 0.015310 & 0.985703 & 0.990788 & 0.974594 \\
ARAur   & 0.952809 & 0.977447 & 1.087226 & 0.197334 & 0.203380 & 0.203380 & 0.000000 & 0.004316 & 0.004316 & 0.000000 & 0.049768 & 0.049768 & 0.999666 & 1.000000 & 0.999998 \\
AYCam   & 1.020000 & 1.033210 & 1.020641 & 0.379918 & 0.396740 & 0.396740 & 0.000000 & 0.000851 & 0.000851 & 0.000000 & 0.032617 & 0.032617 & 0.999641 & 0.998425 & 0.997231 \\
BPVul   & 0.878843 & 0.794209 & n/a      & 0.351141 & 0.347368 & n/a      & 0.001672 & 0.001697 & n/a      & 0.035461 & 0.048946 & n/a      & 0.999173 & 0.998781 & n/a      \\
BSDra   & 0.998519 & 1.015986 & n/a      & 0.222700 & 0.235498 & n/a      & 0.000000 & 0.002003 & n/a      & 0.000000 & 0.023227 & n/a      & 0.999945 & 1.000000 & n/a      \\
CDTau   & 1.000000 & 1.013542 & 1.013542 & 0.250284 & 0.255794 & 0.255794 & 0.000000 & 0.000038 & 0.000038 & 0.000000 & 0.003531 & 0.003531 & 0.999194 & 0.999689 & 0.998063 \\
CVBoo   & 0.974931 & 0.984410 & 0.945464 & 0.537189 & 0.528596 & 0.528596 & 0.000000 & 0.001068 & 0.001068 & 0.000000 & 0.013980 & 0.013980 & 0.999322 & 1.000000 & 1.000000 \\
CWCMa   & 0.953704 & 0.980575 & n/a      & 0.331275 & 0.323462 & n/a      & 0.000000 & 0.000571 & n/a      & 0.000000 & 0.006761 & n/a      & 0.993171 & 0.994072 & n/a      \\
DIHer   & 0.888235 & 0.886187 & 0.886187 & 0.126978 & 0.131158 & 0.135268 & 0.048925 & 0.052926 & 0.057922 & 0.486546 & 0.530156 & 0.613149 & 0.999925 & 1.000000 & 1.000000 \\
EKCep   & 0.632462 & 0.520058 & 0.407451 & 0.176199 & 0.153348 & 0.153348 & 0.001653 & 0.001314 & 0.001314 & 0.108987 & 0.104116 & 0.104116 & 0.999893 & 1.000000 & 0.999962 \\
EOVel   & 1.000000 & 1.006495 & 1.006495 & 0.287392 & 0.290418 & 0.290418 & 0.019802 & 0.018724 & 0.018724 & 0.207055 & 0.190249 & 0.190249 & 0.998537 & 0.999836 & 0.996450 \\
EWOri   & 0.968342 & 0.916530 & 0.916530 & 0.110157 & 0.092226 & 0.091502 & 0.006494 & 0.000175 & 0.001745 & 0.067689 & 0.220550 & 0.099985 & 0.999981 & 1.000000 & 1.000000 \\
FLLyr   & 0.861021 & 0.726411 & 0.636784 & 0.197047 & 0.192678 & 0.192678 & 0.000000 & 0.001047 & 0.001047 & 0.000000 & 0.041959 & 0.041959 & 0.997916 & 0.998066 & 0.998066 \\
FSMon   & 0.975428 & 0.956278 & n/a      & 0.391090 & 0.384414 & n/a      & 0.000000 & 0.000461 & n/a      & 0.000000 & 0.015562 & n/a      & 0.999194 & 0.998205 & n/a      \\
GGOri   & 1.000000 & 0.962852 & 0.962852 & 0.143522 & 0.125972 & 0.125972 & 0.008292 & 0.007391 & 0.007391 & 0.221645 & 0.184309 & 0.184309 & 0.999912 & 1.000000 & 1.000000 \\
ADAnd   & 1.000000 & 1.000111 & n/a      & 0.635524 & 0.611354 & n/a      & 0.000000 & 0.006577 & n/a      & 0.000000 & 0.079439 & n/a      & 0.990024 & 0.990234 & n/a      \\
KPAql   & 1.000000 & 1.008796 & n/a      & 0.262120 & 0.268698 & n/a      & 0.000000 & 0.001840 & n/a      & 0.000000 & 0.064800 & n/a      & 1.000000 & 0.999135 & n/a      \\
MNCas   & 0.981771 & 0.981354 & 0.981354 & 0.643494 & 0.610858 & 0.610860 & 0.000000 & 0.003552 & 0.003552 & 0.000000 & 0.043670 & 0.043670 & 0.982935 & 0.993720 & 0.982159 \\
MUCas   & 0.976821 & 0.974629 & 0.974629 & 0.201461 & 0.193944 & 0.193944 & 0.000788 & 0.000391 & 0.000391 & 0.192998 & 0.197634 & 0.197634 & 0.998648 & 1.000000 & 0.997711 \\
RTCrB   & 0.869711 & 0.718006 & n/a      & 0.291449 & 0.291368 & n/a      & 0.000000 & 0.001057 & n/a      & 0.000000 & 0.029710 & n/a      & 0.997053 & 0.996388 & n/a      \\
HSAur   & 0.972690 & 0.863995 & 0.863995 & 0.080000 & 0.081268 & 0.081268 & 0.000000 & 0.001958 & 0.002247 & 0.000000 & 0.084165 & 0.084158 & 0.999986 & 0.999884 & 0.999884 \\
AOMon   & 0.939669 & 0.953828 & n/a      & 0.475458 & 0.493134 & n/a      & 0.004689 & 0.001165 & n/a      & 0.058813 & 0.015671 & n/a      & 0.999263 & 0.998298 & n/a      \\
ZZBoo   & 1.000000 & 1.021490 & n/a      & 0.238124 & 0.235394 & n/a      & 0.000000 & 0.004847 & n/a      & 0.000000 & 0.057173 & n/a      & 0.999610 & 1.000000 & n/a      \\
ZHer    & 0.729483 & 0.549924 & 0.463213 & 0.300329 & 0.297174 & 0.297174 & 0.000000 & 0.000463 & 0.000463 & 0.000000 & 0.011840 & 0.011840 & 0.993572 & 0.997686 & 0.995996 \\
YYSgr   & 0.955037 & 0.867063 & n/a      & 0.312906 & 0.298538 & n/a      & 0.010289 & 0.007469 & n/a      & 0.157164 & 0.125959 & n/a      & 0.999812 & 1.000000 & n/a      \\
ASCam   & 0.856783 & 0.803458 & 0.803458 & 0.265351 & 0.292222 & 0.292222 & 0.011357 & 0.007212 & 0.007212 & 0.158794 & 0.117928 & 0.117928 & 0.998873 & 0.997001 & 0.997001 \\
WZOph   & 1.003226 & 1.019970 & n/a      & 0.189427 & 0.197384 & n/a      & 0.000000 & 0.003889 & n/a      & 0.000000 & 0.045164 & n/a      & 0.999848 & 1.000000 & n/a      \\
WYHya   & 0.998750 & 0.982007 & n/a      & 0.625098 & 0.593328 & n/a      & 0.000000 & 0.008994 & n/a      & 0.000000 & 0.108341 & n/a      & 0.999560 & 1.000000 & n/a      \\
WXCep   & 0.911111 & 0.864029 & n/a      & 0.377367 & 0.394864 & n/a      & 0.000000 & 0.000327 & n/a      & 0.000000 & 0.004472 & n/a      & 0.996709 & 0.996146 & n/a      \\
WGru    & 1.000000 & 1.016632 & n/a      & 0.340011 & 0.341364 & n/a      & 0.000000 & 0.001065 & n/a      & 0.000000 & 0.040557 & n/a      & 0.998342 & 0.997495 & n/a      \\
ASCam   & 0.955039 & 0.928063 & 1.193993 & 0.218655 & 0.218274 & 0.218274 & 0.000000 & 0.001193 & 0.001193 & 0.000000 & 0.013470 & 0.013470 & 0.999848 & 0.999770 & 0.999770 \\
IQPer   & 0.658537 & 0.598996 & 0.546525 & 0.369499 & 0.373758 & 0.375770 & 0.001599 & 0.000191 & 0.000507 & 0.074983 & 0.009111 & 0.036184 & 0.999925 & 0.992499 & 0.989035 \\
V909Cyg & 0.905719 & 0.995887 & 0.751747 & 0.249693 & 0.195172 & 0.193984 & 0.000000 & 0.012365 & 0.008078 & 0.000000 & 0.143988 & 0.095696 & 0.999903 & 1.000000 & 0.997954 \\
V624Her & 0.974233 & 0.995996 & 0.736358 & 0.311507 & 0.347814 & 0.347814 & 0.000000 & 0.011249 & 0.011249 & 0.000000 & 0.131184 & 0.131184 & 0.982935 & 0.963935 & 0.976785 \\
V526Sgr & 0.836634 & 0.639447 & 0.639447 & 0.327805 & 0.366604 & 0.366436 & 0.017009 & 0.012257 & 0.011699 & 0.218740 & 0.156340 & 0.154611 & 0.998890 & 0.998582 & 0.995845 \\
V523Sgr & 1.000000 & 1.006317 & 1.004570 & 0.379744 & 0.341982 & 0.343246 & 0.003251 & 0.008262 & 0.001237 & 0.161967 & 0.076667 & 0.094935 & 0.992757 & 0.997308 & 0.993745 \\
V541Cyg & 1.000000 & 0.783731 & n/a      & 0.085835 & 0.105824 & n/a      & 0.038308 & 0.009920 & n/a      & 0.477466 & 0.136967 & n/a      & 0.999998 & 1.000000 & n/a      \\
V499Sco & 0.929231 & 0.950073 & 0.948501 & 0.592589 & 0.616102 & 0.634779 & 0.000000 & 0.000751 & 0.000000 & 0.000000 & 0.010311 & 0.000000 & 0.996041 & 0.990092 & 0.983366 \\
V478Cyg & 0.980841 & 1.023842 & 1.023842 & 0.542721 & 0.515278 & 0.515278 & 0.000584 & 0.000670 & 0.000670 & 0.029494 & 0.006192 & 0.006192 & 0.978401 & 0.985782 & 0.980823 \\
V477Cyg & 0.748023 & 0.584170 & 0.581124 & 0.279444 & 0.212522 & 0.218526 & 0.016408 & 0.010979 & 0.019542 & 0.330593 & 0.184558 & 0.325932 & 0.997133 & 1.000000 & 0.998558 \\
V459Cas & 0.995405 & 0.955418 & n/a      & 0.143540 & 0.151228 & n/a      & 0.001776 & 0.002711 & n/a      & 0.024235 & 0.088617 & n/a      & 0.999956 & 1.000000 & n/a      \\
V442Cyg & 0.974638 & 0.988779 & n/a      & 0.343847 & 0.339708 & n/a      & 0.000000 & 0.001121 & n/a      & 0.000000 & 0.013263 & n/a      & 0.997564 & 0.997072 & n/a      \\
V396Cas & 0.926837 & 0.870983 & n/a      & 0.204625 & 0.203678 & n/a      & 0.000000 & 0.000849 & n/a      & 0.000000 & 0.032381 & n/a      & 0.999788 & 0.998848 & n/a      \\
V364Lac & 0.970706 & 1.043457 & 1.043457 & 0.233226 & 0.193654 & 0.196142 & 0.007461 & 0.001641 & 0.005036 & 0.287203 & 0.103638 & 0.197713 & 0.999900 & 1.000000 & 1.000000 \\
V364Cas & 1.000000 & 1.015927 & n/a      & 0.415951 & 0.417392 & n/a      & 0.000000 & 0.003001 & n/a      & 0.000000 & 0.036561 & n/a      & 0.999848 & 1.000000 & n/a      \\
UZDra   & 0.958033 & 0.920909 & 0.825579 & 0.194092 & 0.200270 & 0.200270 & 0.000000 & 0.000848 & 0.000848 & 0.000000 & 0.032346 & 0.032346 & 0.999930 & 1.000000 & 1.000000 \\
SWCMa   & 1.000000 & 0.928059 & 0.928059 & 0.171623 & 0.174596 & 0.176710 & 0.015720 & 0.007733 & 0.015278 & 0.317511 & 0.243566 & 0.314302 & 0.999750 & 0.994065 & 0.999385 \\
STCen   & 0.993076 & 0.981704 & n/a      & 0.646668 & 0.614678 & n/a      & 0.000000 & 0.005525 & n/a      & 0.000000 & 0.066824 & n/a      & 0.993961 & 0.993632 & n/a      \\
NYCep   & 0.798611 & 0.847391 & n/a      & 0.173402 & 0.115962 & n/a      & 0.008480 & 0.002560 & n/a      & 0.479925 & 0.122185 & n/a      & 0.975879 & 0.997099 & n/a      \\
\enddata
\end{deluxetable}

The results for 50 CALEB stars seem promising: the network managed to get a satisfactory starting point for all LCs, and DC converged successfully to the final solution. Comparing ANN results against published solutions cannot be done rigorously since, as stated previously, there are systematic discrepancies in the published data; inspecting Table \ref{caleb_table} and LCs given in Appendix \ref{caleb_plots} reassures that the network, powered by DC, was capable of getting probable parameters for 50 out of 50 CALEB test stars.

For the continuation of this test we initiated an undergraduate project at Villanova and Eastern Universities to review, enhance, and extend the CALEB database with modern data. This ambitious project aims to collect LC and RV curves, spectra, and model solutions for the total of $\sim$1000 EBs. That will be roughly a factor of 3 of improvement when compared to CALEB. This database will be publicly available.

\subsection{Application: 2580 stars from OGLE detached EB catalog}

\begin{figure}
\begin{center}
\includegraphics[width=\textwidth]{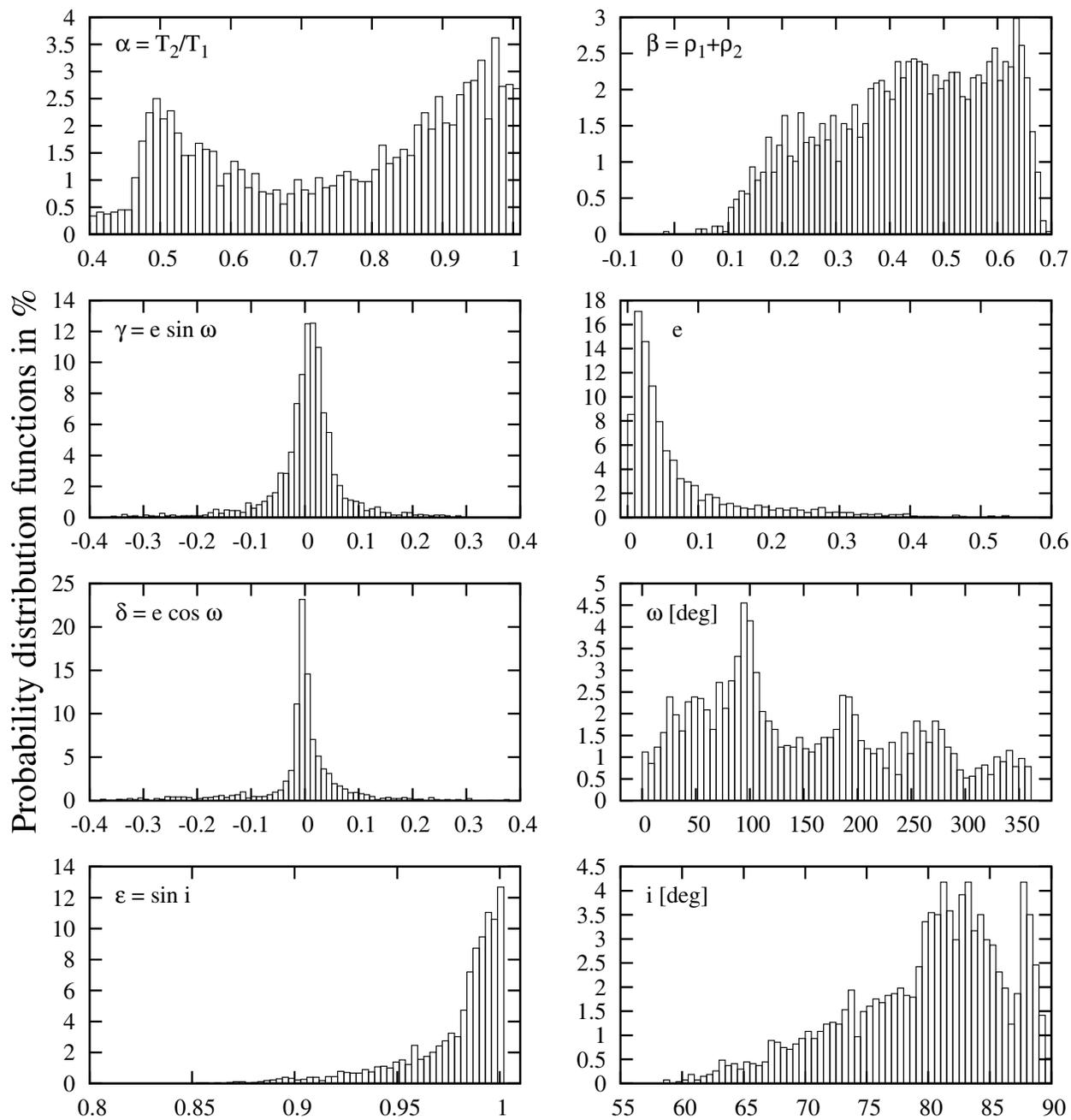} \\
\end{center}
\caption{The distributions of parameters obtained by the ANN from the OGLE sample of 2580 LMC binaries classified as detached.} \label{oglehist}
\end{figure}

\citet{wyrzykowski2003} published a catalog of 2580 detached EBs detected in a 4.6 square degree area of the central parts of the Large Magellanic Cloud (LMC) which includes photometric data collected during the OGLE-2 microlensing search from 1997 to 2000. The sample consists of 2681 entries due to field overlap. The sample was previously used for statistical studies of EBs, e.g.~by \citet{michalska2005}, \citet{mazeh2006} and \citet{faccioli2008}. We took the sample in its entirety, without pre-filtering (no false positive removal, no whitening). \emph{Polyfit} was applied to obtain analytic approximations to light curves and compute equidistant data points for input to the trained ANN. The ANN processing time of the OGLE sample on a 2GHz processor was under 1 second. Fig.~\ref{oglehist} shows the statistical distribution of the obtained parameters by ANN.

The following properties of the OGLE sample are observed:

\begin{itemize}
\item the distribution of $\beta = \rho_1 + \rho_2$ shows a selection effect in the sample: the number of data points per light curve, coupled with the classification scheme used to generate the model (namely image recognition on a rasterized/degraded light curve -- see \citet{wyrzykowski2003} for details), favors close contact binaries;
\item the distributions of eccentricity $e$ and argument of periastron $\omega$ are coupled. There is an apparent deficiency of systems with near zero eccentricity (only 8\% instead of the expected $\sim$30\%) on account of the excess of low eccentricity systems with arguments of periastron close to $90^\circ$ and $270^\circ$. In these cases the line of sight is aligned with the semi-major axis and the eclipses are exactly half a period apart, and the sum of excesses exactly matches the deficiency in zero eccentricities. This is \emph{not a systematic effect}; rather, the inverse problem is \emph{inherently degenerate} -- there is no information in the light curves of detached EBs that would allow discrimination between these systems.
\item the distribution of $\epsilon = \sin i$ closely follows the expected distribution, with the exception of a slight deficiency of systems close to $i=87^\circ$. Given that higher inclinations typically correspond to total eclipses, and smaller inclinations correspond to partial eclipses, there is no solid handle for systems close to this boundary: these can be either total or partial eclipses and the network cannot discriminate. This effect is again \emph{inherent} to the problem, this time the degeneracy is between the inclination and radii (see \citet{phoebe} for a thorough discussion of this degeneracy).
\item the most interesting distribution is the temperature ratio $\alpha = T_2/T_1$. There are two obvious contributions: a half Gaussian-like distribution with the center around $T_2/T_1 = 1$, and a distinct excess superimposed at approximately $T_2/T_1 = 0.5$. Careful analysis has shown that the $T_2/T_1$ excess has an exact counterpart in the $\beta$ distribution, between 0.4 and 0.5. There is no correlation observed between $T_2/T_1$ and eccentricity (which could in principle exist because of the influence of eccentricity on eclipse depth), nor in fact with any other attained quantity except $\rho_1+\rho_2$. We also ascertained that there is no similar excess in a synthetic sample. The source of the hump around $T_2/T_1 = 0.5$ is not entirely clear; it could be due either to the selection effect of the OGLE sample (i.e.~semi-detached and overcontact EB contamination of the sample), or be caused by a real physical duality in the sample. Either way this result merits further study.
\end{itemize}

The obtained parameters are not final. The next step would be to feed the ANN results into the light curve solver, similar to what was done for the CALEB data presented in Section \ref{caleb_main}. Since this paper focuses on the performance of the ANN tier alone, we defer a more thorough analysis of the OGLE data to a follow-up paper.

The results, although rudimentary and uncorrected for selection effects, allow comparison with the extensive study of LMC short period binaries by \citet{mazeh2006}. They draw their sample from the same pool of 2580 EBs, trimmed by a detectability and quality factor. Their data were limited to: periods less than 10 days, $I$-band magnitudes between 17 and 19; binary components on or close to the main sequence, yielding a sample of 938 binaries. Our histogram of the sum of radii (cf.~\ref{oglehist}; top right) matches \citeauthor{mazeh2006}'s result well and generalizes it to a certain extent because of the unfiltered sample and the underlying Roche formulation, as it properly accounts for distorted stars. We also find the bimodal histogram in surface brightness (in our case its proxy $T_2/T_1$; top left), which \citeauthor{mazeh2006} attribute to short period, presumably mass-transferred binaries. Note that this feature is present even after their selection correction. \citeauthor{mazeh2006} do not use the inclinations in their statistical analysis. As noted earlier, our inclination distribution (bottom right inset) is somewhat hindered by the degeneracy, but its overall statistical significance should be valid for the whole OGLE LMC sample of EBs.

\subsubsection{Inferential science}

EB parameters gathered from large surveys can yield statistical insight into the physics governing their distributions. Since our tests of ANNs are shown to provide statistically viable results, the question is what scientific results can we extract from the sample. Even in the absence of thorough analysis with heuristic error estimates, additional data and dedicated follow-ups, the science extracted from large numbers of automatically analyzed LCs can nevertheless be important because we learn physical properties by inference from the whole sample.

\begin{figure}
\begin{center}
\includegraphics[height=0.46\textwidth,angle=-90]{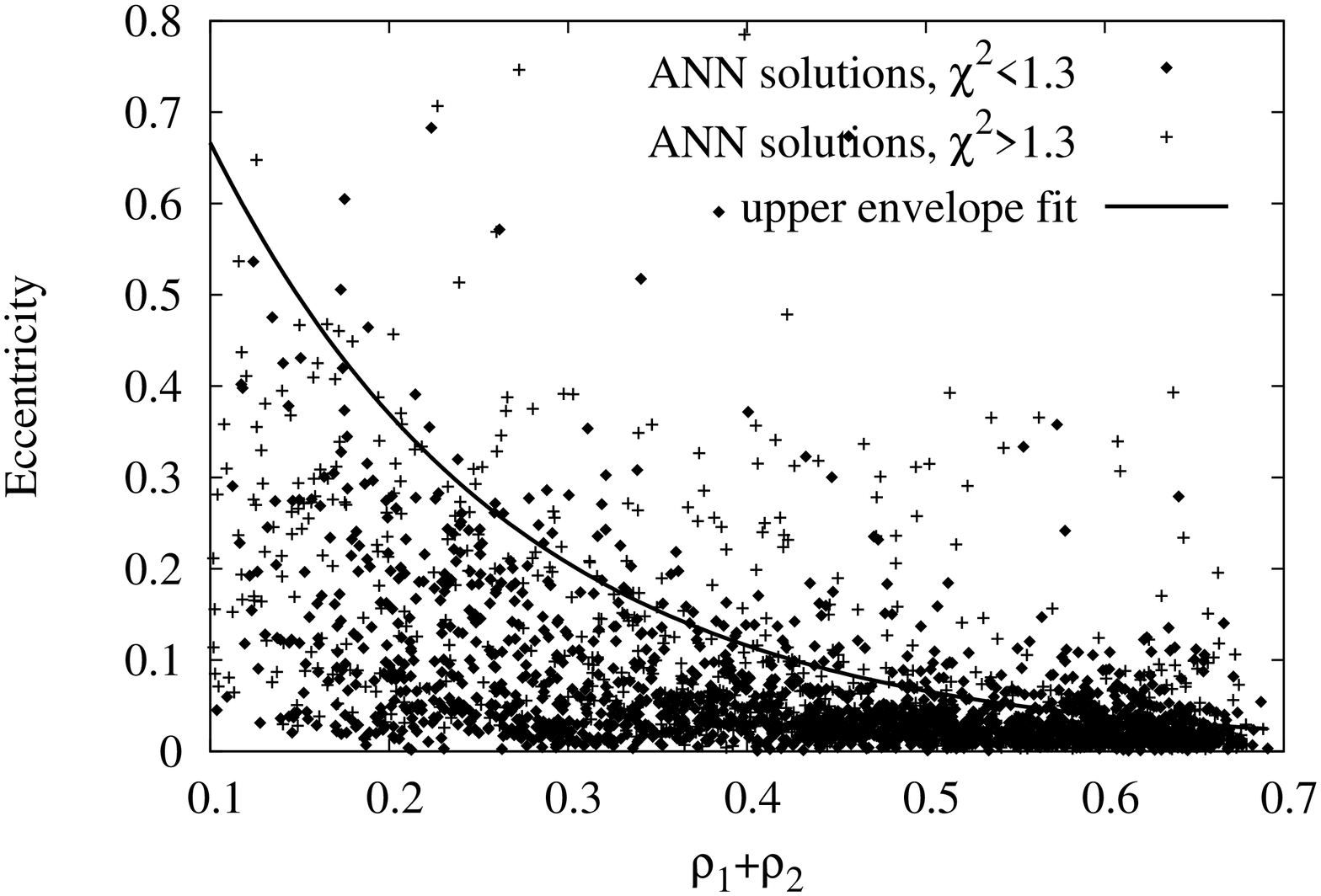}
\includegraphics[height=0.44\textwidth,angle=-90]{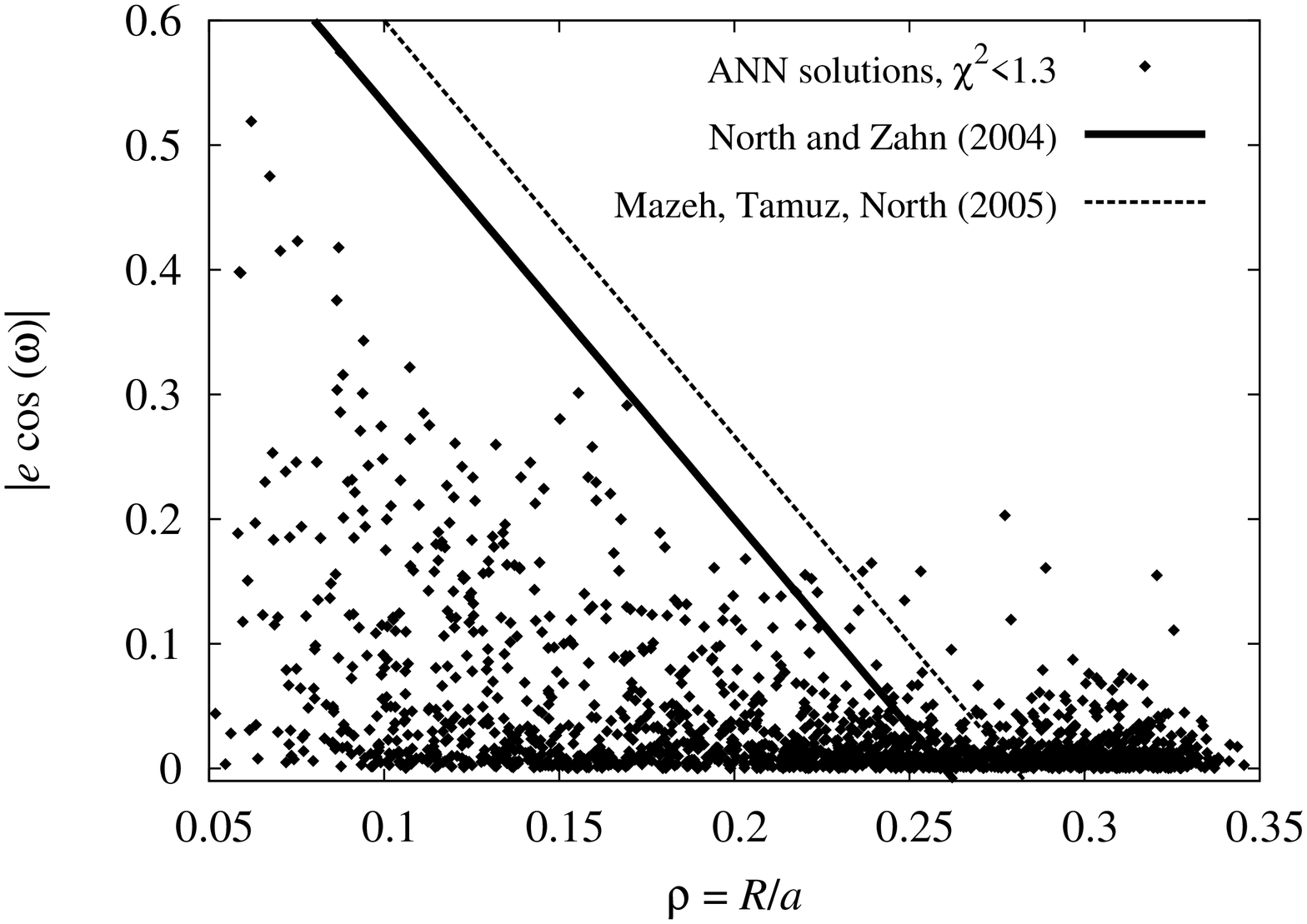} \\
\end{center}
\caption{Left: The correlation between the sum of fractional radii and eccentricity for the OGLE LMC sample of 2681 stars. Diamonds denote solutions with $\chi^2$ smaller than 1.3, and pluses denote solutions with $\chi^2$ larger than 1.3. The values of $\chi^2$ are computed by feeding the ANN-derived parameters to \phoebents, computing the corresponding synthetic curve, scaling it vertically to match the input curve and computing the residuals. The solid line represents the exponential function fit to the $3\sigma$-$1\sigma$ boundary, where all points were taken into account. The unusually large number of poor fits above the boundary is due to close-to-contact and overcontact EBs in the OGLE sample which the network was not trained to recognize. The increased number of poor fits towards low $\rho_1+\rho_2$ is due to training set under-sampling and phase resolution deficiency, as discussed in Section \ref{nettraining}.--- Right: $|e \cos \omega|$ versus relative radius for all OGLE LMC binaries with $\chi^2$ smaller than 1.3. A solid line represents the solution from \citet{north2004} derived from a sub-sample of B stars; a dashed line represents the solution from \citet{mazeh2006}. The critical radius where orbits rapidly become circular is $\rho_c \sim 0.26$, in perfect agreement with the previous studies.} \label{betaecc}
\end{figure}

An example of inferential science is the correlation between the sum of fractional radii and eccentricity. It is well known that circularization and synchronization mechanisms depend strongly on the relative separation between stars due to tidal forces, torque, magnetic coupling, etc, leading to a correlation of $\rho_1+\rho_2$ with $e$. Fig.~\ref{betaecc} (left) shows the results for the OGLE sample: for each binary, EB parameters yielded by the network were fed back to the modeling engine, a synthetic LC computed and passband luminosity adjusted to match the data. The $\chi^2$ value, computed from the residuals was used to qualify the fit: diamonds in Fig.~\ref{betaecc} represent solutions with $\chi^2 \leq 1.3$, whereas pluses represent solutions with $\chi^2 > 1.3$. The solid line is an exponential function fit to the data so that points with up to $3\sigma$ deviation lie below and points with up to $1\sigma$ deviation lie above the boundary. The fits are systematically poorer at lower values of $\rho_1+\rho_2$ due to under-sampling in very narrow eclipses and the under-sampling of the network. Fit degradation can also be observed above the boundary due to contamination of the detached OGLE database with close-to-contact and overcontact systems that the network was not trained to recognize. Fig.~\ref{betaecc} reveals a strong correlation between $\rho_1+\rho_2$ and $e$. Tidal circularization would suggest that stars close to the boundary are young, with age increasing gradually toward the bottom of the plot. The exponential-like dependence is intriguing and warrants comparison with a more rigorous model to derive specific results for circularization rates. \citet{north2004} and \citet{mazeh2006} observed the dependence of $|e \cos \omega|$ on the relative radius $\rho$; the equivalent plot for ANN data with $\chi^2 < 1.3$ is depicted on Fig.~\ref{betaecc} (right). The results of \citet{north2004}, performed on a filtered sample of nearly equal B-type components, indicate a linear dependence of $|e \cos \omega|$ on $\rho$, with a sharp cut-off $\rho_C$ at the onset of which all orbits appear circularized. The solid line in Fig.~\ref{betaecc} represents the occurence boundary observed by \citet{north2004} and a dashed line represents the boundary observed by \citet{mazeh2006}. Our data show remarkable agreement in the estimate of $\rho_C$, but possibly suggest a somewhat shallower linear dependence than that presented by the previous studies. This might be a consequence of approximating $\langle \rho \rangle = \beta/2$ for the whole OGLE sample and will be studied more thoroughly in future.

\subsection{Discussion}

The EBAI project is motivated by the goal of maximizing the scientific output resulting from the analysis of the future fire-hose of binary star data.  For this, we are committed to the application of a rigorous physical model (here, the WD model) that encompasses the complete range of binary star configurations.  Moreover, we want no efficiency penalty for this, and ANNs are a practical means to this end, capable of performing a forward pass of thousands of observed light curves through the trained network in a fraction of the time required to compute a single theoretical light curve!  While this paper focuses on detached EBs, the method is readily extendable to systems with large proximity effects as fully treated in the WD model and will be discussed in a subsequent paper.  We will thus be able to apply ANNs in a unified, automated approach to solutions for eclipsing binary light curves of all classes. By contrast, \citet{devor2005}'s solution approach is time efficient, but limited to well-detached systems with no proximity effects.  \citet{tamuz2006}'s approach employs a more realistic EBOP model, which is still limited in handling proximity physics.  Nevertheless, these pioneering efforts have indeed shown their value in applications to large survey data.  A computational penalty, if any, for using more sophisticated models will eventually vanish, thanks to Moore's Law.

The benefits of ANNs here are not limited to speed, but include a non-linear interpolation capability widely claimed to be robust to noise \citep[e.g.~][pg.~246]{freeman1991}.  Characteristically for ANNs such a property is difficult to quantify, but our experience supports this claim. Preliminary tests have been done on light curves with significant secular variations, most notably due to chromospheric activity and proximity effects. One such example is V380 Cyg \citep{guinan2000}, an active eccentric binary with total eclipses. Although the feedback from the network failed to reproduce the shape of the light curve (as there is no provision for such secular variations), the parameters are reasonably close to the published values. This observation indicates that the ANN could cope with such challenging cases even though the light curve is severely affected by the secular effects. We plan to test such EBs further and to assess the sensitivity of the ANN on the amplitude and periods of intrinsic variations.

The output of ANN is not a final solution; rather, it provides initial parameter estimates for dedicated EB modeling engines. These engines refine the result, estimate formal and heuristic errors of the fit, and produce statistics for the sample. A preview of this approach has been presented for CALEB data in Section \ref{caleb_main}.

But what happens when the EB-trained ANN is input some non-EB periodic light curve (a Cepheid, RR Lyr, etc.)?  By itself, the ANN tier has no way of detecting this, it will produce a set of parameters for such a curve as well. However, the very next step in the IDP is to pass these parameters to \phoebe and to compute the $\chi^2$ value of the solution. This value is an immediate indicator of the goodness-of-fit and can be used for rejection, re-training of the classifier, and subsequent re-classification. This procedure is expected to be quite effective, since the WD model acts as a \emph{matched filter} for EBs, and this property is inherited by the EB-trained ANN.  

\section{Conclusion}

ANNs have proved successful in problems of classification, real-time control, data mining and many other tasks in a variety of scientific and technical applications. This paper demonstrates their utility in parameter estimation for detached binaries, even in the face of parameter degeneracy, the correlation of certain pairs of parameters; e.g., the compensating effect of increasing a system's sum of radii vs. increasing its inclination.

A task-optimal ANN topology and learning rate was determined and the network was trained on the rigorous WD model.  The trained network was tested on a previously unseen test set of 10,000 WD model light curves, yielding model parameters to better than 90\% accuracy for 90\% of the test set.  In tests with 50 real-world detached binary systems from the CALEB database, the ANN alone achieved sufficiently good fits for 22 systems that differential corrections were not needed.  Those requiring differential corrections all converged to produce good quality fits, confirming the utility of the ANN-produced starting parameters.  Overall, the CALEB test produced a 100\% success rate for the ANN.

We have shown that a suitable ANN can be successfully trained on a sophisticated EB model, and that the trained network produces quite satisfactory approximate light curve solutions with high computational efficiency.  In addition to these advantages, ANNs have favorable properties of interpolation; they are well-behaved in regions around their multi-dimensional training points.  Thus the network will interpolate reasonable starting parameters for light curves with asymmetry or spots, and in fact for any (detached) light curve that it has not seen before.  It is quite remarkable that the network is able to memorize to such high quality all the light curve solutions in the space of detached binaries with only 8240 numbers ($201\times40 + 40 \times 5$ weights).  Clearly, this network has sufficient information storage capacity, but it is worth noting that the information capacity corresponding to a given network topology is only known in heuristic approximation \citep{neelakanta1994}. This confirms that much is yet to be learned about ANNs, and in particular it explains the necessity to explore network topologies and learning rates for their appropriateness to the given problem.  Such exploration, in addition to the ability to rapidly train many iterations, further emphasizes the need for speed in the learning phase.

This work was greatly facilitated by two developments.  The \emph{polyfit} algorithm produced excellent analytic approximations to real-world light curves sampled at (often quite) unequal phase intervals, where splines or other interpolating methods behave badly \citep{emery2001}, making equal-phase interpolated data points available for input to the ANN.  This algorithm has been found to be an excellent interpolator for EBs of all classes, as well as, for example, Cepheid variables.  Since well-behaved interpolation is a problem common to many fields, \emph{polyfit} could well see further application beyond astronomy.  The scheme for parallelization of ANN training over multiple processors was key to quickly exploring topologies and training parameters, and churning through a half-million training iterations in days instead of weeks.  This scheme had the beneficial side effect of lending simulated annealing behavior to the training phase, improving the search for the global minimum.

The source code for \emph{polyfit} and ANN is released under the GNU General Public License, which grants users the right to freely use, distribute and modify the code. The programs may be downloaded from the IDP project homepage: {\tt http://www.eclipsingbinaries.org} or \phoebe homepage: {\tt http://phoebe.fiz.uni-lj.si}.

This paper focuses on detached EBs.  Systems with large proximity effects which are fully treated in the WD model will be discussed in a subsequent paper.  We will thus be able to apply ANNs in a unified, automated approach to provide starting solutions for eclipsing binary light curves of all classes.

\acknowledgements
This research is supported by NSF/RUI Grant No. AST-05-07542, which we gratefully acknowledge. Computations are performed on a dedicated 24-node Beowulf cluster financed in part by the Astronomy Department of Villanova University; without it, extensive \ebai testing would not be possible. Our thanks go to Dr.~M.~Horvat from University of Ljubljana for useful comments on the polyfit implementation, and to Dr.~T.~Zwitter for his constructive comments on the manuscript clarity. We are indebted to the anonymous referee for thorough reading and excellent suggestions that improved the quality of the paper.

\appendix

\section{Polyfit algorithm - details} \label{polyfit_appendix}

In Section \ref{polyfit_section} we introduced a light curve fitting algorithm \emph{polyfit}: a polynomial chain fitter. In this appendix we provide a formal description and comment on the implementation details. The \emph{polyfit} algorithm fits a polynomial chain $\mathcal P(x)$ of piecewise smooth $n$-th order polynomials connected at the given set of knots. The two key ideas of the algorithm are to abandon the requirement of differentiability of $\mathcal P(x)$ at knots and thus allow the polynomial chain to be broken, and to randomly perturb the position of the knots and relax the system to the nearest minimum.

\begin{enumerate}
\item Given the set of knots $x_k$; $k = 1 \dots N$, $x_{k+1} > x_k$ $\forall k$, partition the phase range into $N$ intervals:
$$ I_1 = [x_1, x_2), \quad I_2 = [x_2, x_3), \quad \dots,\quad I_N = [x_N, x_1).$$
\item Given the first interval $I_1 = [x_1, x_2)$, do an $n$-th order polynomial least squares fit through data points on that interval:
$$ p_1 (x) = a^1_n (x-x_1)^n + a^1_{n-1} (x-x_1)^{n-1} + \dots + a^1_1 (x-x_1) + a^1_0, \quad x \in [x_1, x_2) \equiv I_1. $$
\item Given the set of intervals $I_k$, $k = 2 \dots (N-1)$, do the $n$-th order polynomial least squares fit through the data enclosed between two adjacent knots:
$$ p_k (x) = a^k_n (x-x_k)^n + a^k_{n-1} (x-x_k)^{n-1} + \dots + a^k_1 (x-x_k) + a^k_0, \quad x \in [x_k, x_{k+1}) \equiv I_k $$
while satisfying the constraint that the polynomial $p_k(x)$ must be connected with $p_{k-1}(x)$ at the knot $x_k$:
$$ p_k (x_k) = p_{k-1} (x_k): \quad a_0^k = a^{k-1}_n (x_k-x_{k-1})^n + a^{k-1}_{n-1} (x_k-x_{k-1})^{n-1} + \dots + a^{k-1}_0. $$
The last equation is a recursive formula to compute $a^k_0$ from the coefficients of $p_{k-1} (x)$.

\item Given the last interval $I_N = [x_N, x_1)$, where $x_1$ is aliased ($x_1 \mapsto x_1+1.0$), do the $n$-th order polynomial least squares fit through data points on that interval:
$$ p_N (x) = a^N_n (x-x_N)^n + a^N_{n-1} (x-x_N)^{n-1} + \dots + a^N_1 (x-x_N) + a^N_0, \quad x \in [x_N, x_1) \equiv I_N $$
while satisfying two constraints, connectivity:
$$ p_N (x_N) = p_{N-1} (x_N): \quad a_0^N = a^{N-1}_n (x_N-x_{N-1})^n + a^{N-1}_{n-1} (x_N-x_{N-1})^{n-1} + \dots + a^{N-1}_0 $$
and phase space wrapping:
$$ p_N (x_1) = p_1 (x_1): \quad a_0^1 = a^N_n (x_1-x_N)^n + a^N_{n-1} (x_1-x_N)^{n-1} + \dots + a^N_1 (x_1-x_N) + a^N_0. $$
Since $a_0^1$ has already been determined in step 2, and $a_0^N$ has been determined by the connectivity constraint, this constraint is solved for $a^N_1$ and substitute the resulting expression into $p_N(x)$:
$$ p_N (x) = a_0^N + \sum_{i=2}^n a_i^N (x-x_N) \left[ (x-x_N)^{i-1}-(x_1-x_N)^{i-1} \right].
$$

Least squares fit provides $n+1$ coefficients $a_i^0$ for the first interval, $n$ coefficients $a_i^k$ for intervals 2--($N-1$), and $n-1$ coefficients $a_i^N$ for the last interval, totaling $nN$ coefficients of the \emph{wrapped and connected polynomial chain} $\mathcal P(x)$, $x \in [x_1, x_{N+1} \equiv x_1)$, of piecewise smooth polynomial functions.

\item Given the polynomial chain $\mathcal P(x)$, evaluate the sum of squares of residuals:
$$ \chi^2_1 = \sum_j w_j \left( \mathcal P(x_j) - y_j \right)^2 $$
for all observed data points $(x_j, y_j)$ with respective weights $w_j$.

\item Once the polynomial chain $\mathcal P(x)$ is set and its goodness-of-fit value $\chi^2$ computed, the initial set of knots is varied. Given a step amplitude $\delta$, displace each knot by:
$$ x_k \mapsto x_k + \mathcal N(x_k,\delta), $$
where $\mathcal N(x,\sigma)$ is the Gaussian probability distribution function. Repeat steps 2--5 to compute $\chi^2_2$. If $\chi^2_2 < \chi^2_1$, adopt the new set of knots, else reject it.

\item Repeat steps 2--6 for the given the number of iterations $L$. The final $\chi^2_L$ will correspond to the set of knots $x_k$ that best matches the data.
\end{enumerate}

There are some subtleties of the algorithm that had to be incorporated for our particular application:
\begin{itemize}
\item [a)] there has to be an explicit check that the interval $I_1$ contains at least $n+1$ data points, intervals $I_k$, $k=2\dots(N-1)$, contain at least $n$ data points, and interval $I_N$ contains at least $n-1$ data points.
\item [b)] there has to be a slight repulsion between knots so that the algorithm does not push the knots together. This is most easily achieved by adding a penalty to the cost function of the form $\epsilon (x_k-x_{k-1})^{-2}$ for the given penalizing coefficient $\epsilon$.
\item [c)] the last knot is the same as the first knot: for $N$ knots there are $N$ intervals and not $N-1$ as one may initially expect. The knot on the phase range boundary should thus wrap smoothly over that boundary.
\end{itemize}

\begin{figure}
\includegraphics[angle=-90,width=16cm]{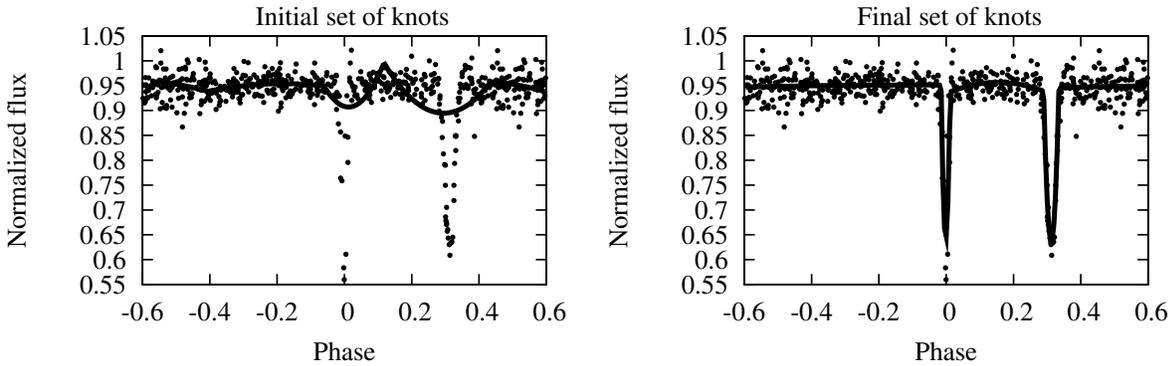} \\
\caption{Performance of the quadratic 4-knot \emph{polyfit} algorithm on an eccentric binary light curve. Left: the solution of a polynomial chain fit with the default set of initial knots, $\{-0.4, -0.1, 0.1, 0.4\}$. Right: the solution of the polynomial chain fit after 5000 iterations.} \label{polyfit}
\end{figure}

For detached EB light curves we found that 4 knots of a quadratic polynomial chain set initially at $\{-0.4, -0.1, 0.1, 0.4\}$ suffice for a 95\% success rate in fitting 2681 OGLE EBs. Actually, 4 knots of a quadratic polynomial chain are likely to suffice for most photometric light curves and can readily be used for classification purposes. The starting set of knots was chosen in the following manner:
\begin{enumerate}
\item the data are sorted in phase;
\item an average flux $\bar f$ of all data points is computed;
\item each\footnote{The first chain starts only after a data point that exceeds the average value $\bar f$ has been encountered; that is done so that the last chain can be wrapped at the phase end.} data point $y_i$ that is \emph{smaller} than $\bar f$ is taken as start of a chain. The chain is built by all data points $y_k$, $k > i$, that are smaller than $\bar f$;
\item if the chain is shorter than a prescribed threshold (we used 10 elements), it is discarded; else it is stored;
\item in case of two or more chains, the two longest chains are picked; the first and last elements of each chain are taken as initial knots. If there are less than 2 chains, fall back on default knots: $\{-0.4, -0.1, 0.1, 0.4\}$.
\end{enumerate}

\begin{figure}
\begin{center}
\includegraphics[angle=-90,width=14cm]{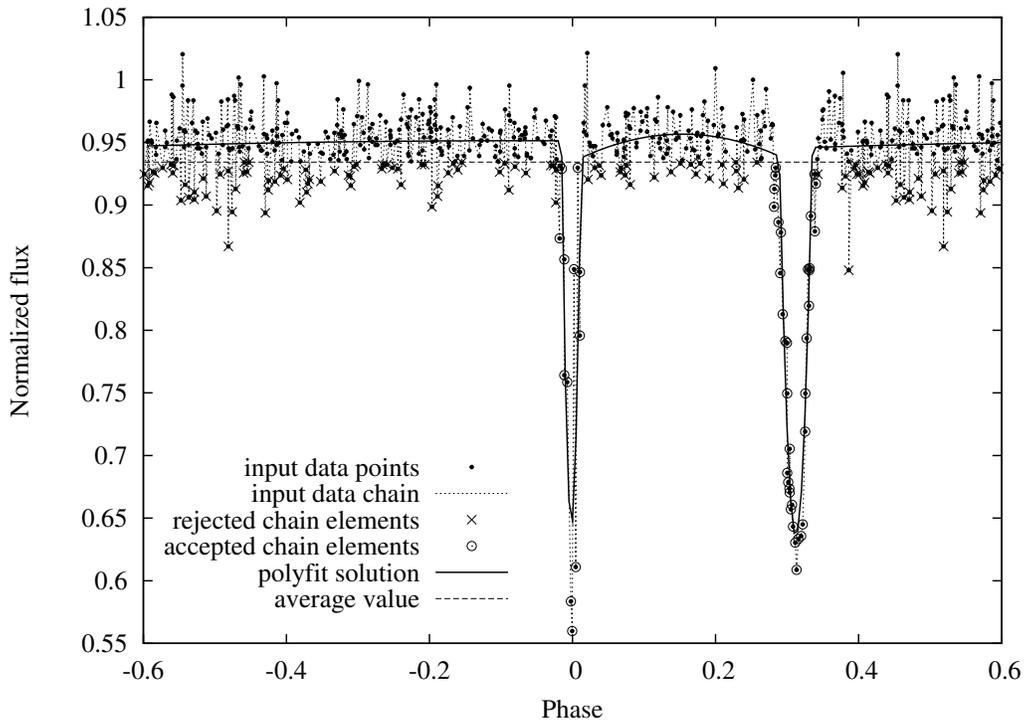} \\
\end{center}
\caption{An example of the automatic knot seeking algorithm performance. This light curve corresponds to a well detached, eccentric binary that proved to be difficult to fit because of the narrow minima and sparse data points with substantial noise covering the primary minimum. Still, \emph{polyfit} performed reasonably well.} \label{autoknots}
\end{figure}

Fig.~\ref{autoknots} shows the difficult example of choosing a set of initial knots by the above algorithm. The required number of iterations was set to $L=5000$. These choices have proved to be particularly robust even in the tough cases of well-detached eccentric binaries (cf.~Fig.~\ref{polyfit}). The algorithm fails in less than 1\% of the tested sample. The time cost of fitting 2681 light curves with $\sim$500 data points each is around 2 minutes on a single 2GHz processor.

\section{Network fits to CALEB light curves} \label{caleb_plots}

In Section \ref{caleb_main} we submitted 50 real-world EBs from the CALEB database to the ANN. Since published model parameters did not always match the data exactly, direct comparison of the results yielded by the network suffers from systematics. As explained in the main part of the paper, parameters obtained from the ANN were used to create a synthetic LC that was normalized to match the data. Based on the computed residuals we submitted the ANN solution to 1, 2, or 3 DC iterations: for $\chi^2$ values of 1.5 or less, no DC iteration was performed and the ANN solution is adopted as final. For $\chi^2$ values between 1.5 and 2.0, 1 DC iteration was performed; for $\chi^2$ values between 2.0 and 2.5, 2 DC iterations were performed; finally, for $\chi^2$ values of 2.5 and higher, 3 DC iterations are performed. Table \ref{caleb_table} lists the systems where DC iterations were necessary; here in the Appendix we overplot solution LCs to the data to demonstrate the success of the method.

\begin{figure}
\caption{EBAI model light curves derived for the CALEB data.} \label{caleb_lcs1}
\begin{center}
\includegraphics[width=\textwidth]{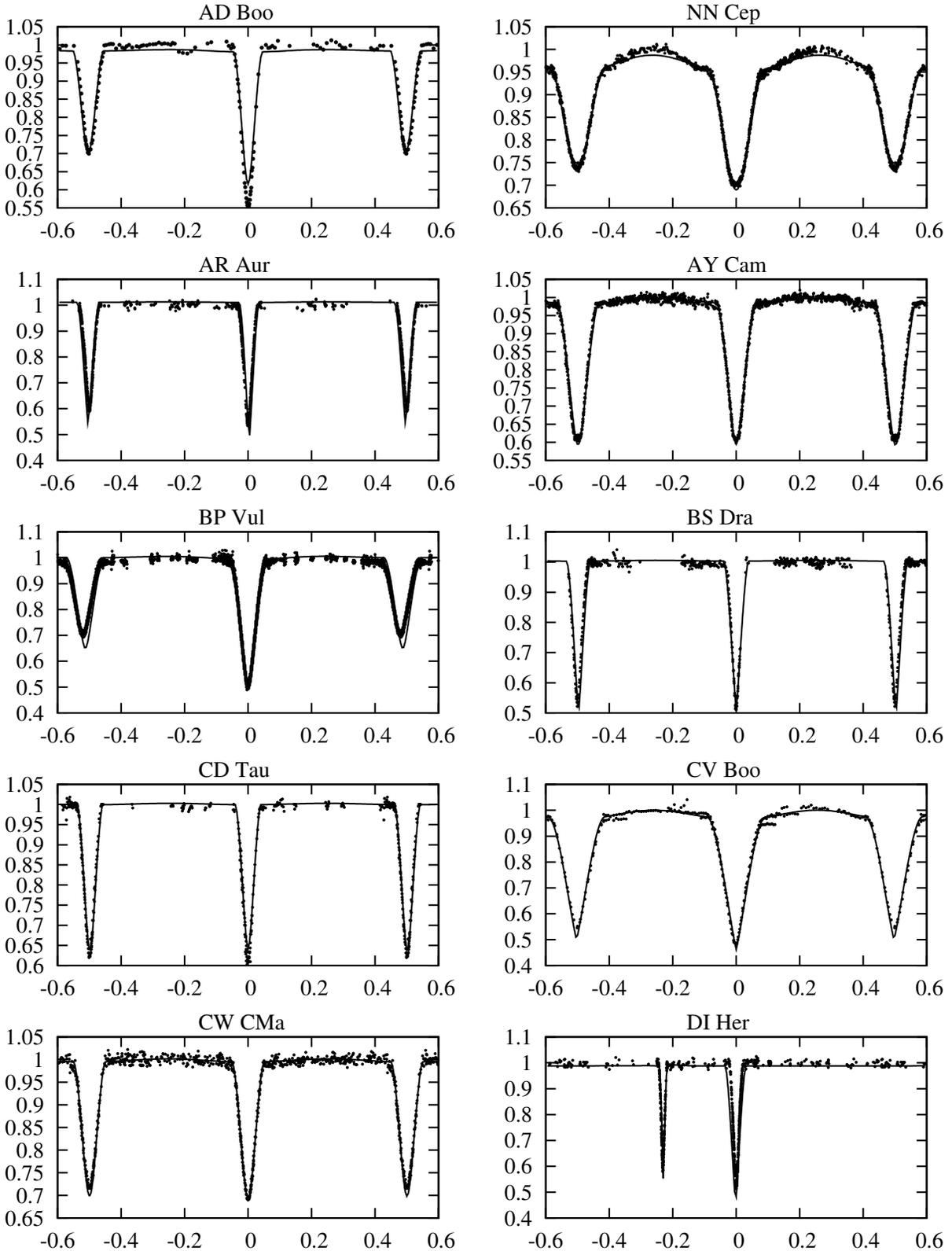} \\
\end{center}
\end{figure}

\begin{figure}
\caption{Continued: EBAI model light curves derived for the CALEB data.} \label{caleb_lcs2}
\begin{center}
\includegraphics[width=\textwidth]{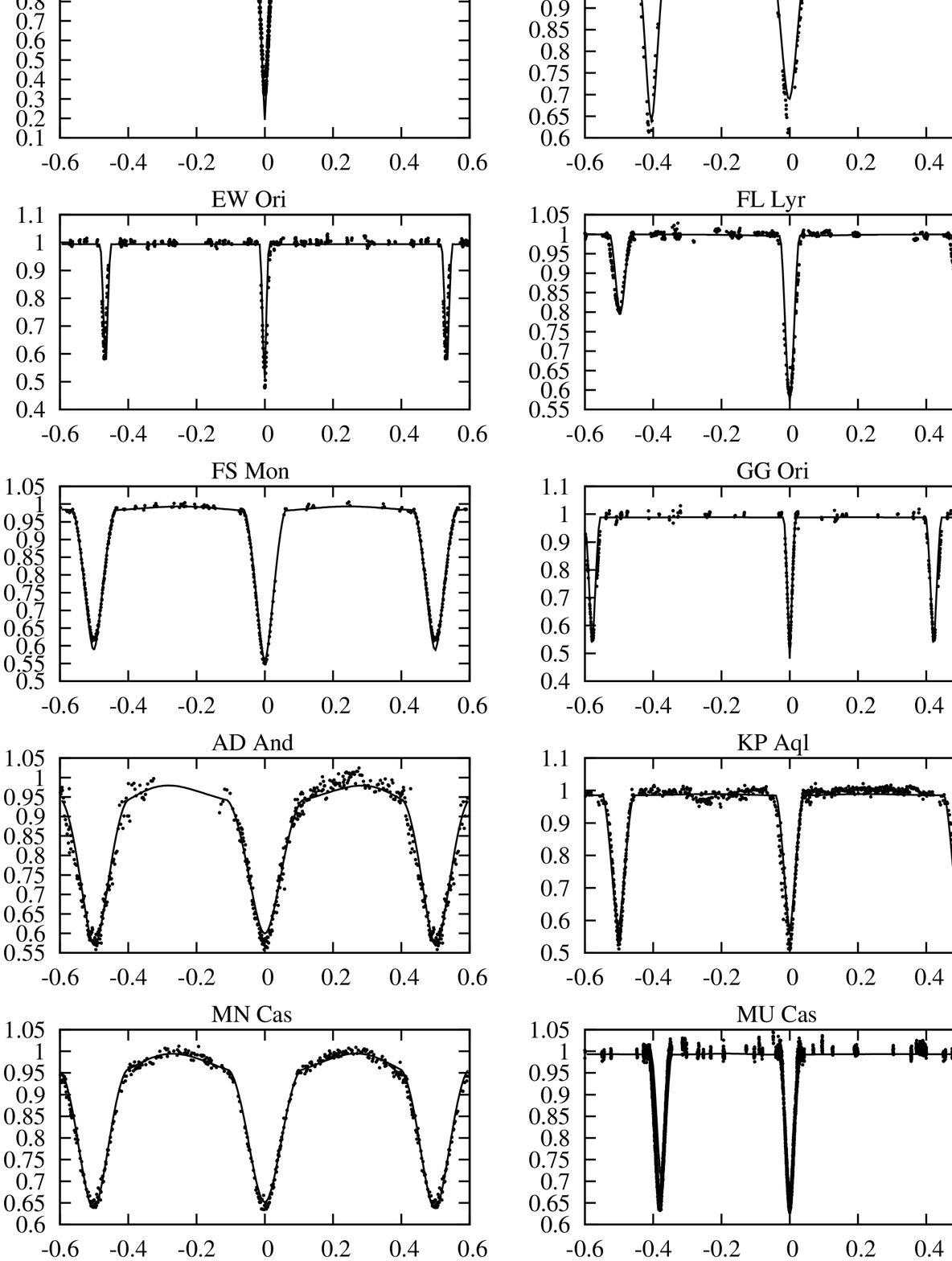} \\
\end{center}
\end{figure}

\begin{figure}
\caption{Continued: EBAI model light curves derived for the CALEB data.} \label{caleb_lcs3}
\begin{center}
\includegraphics[width=\textwidth]{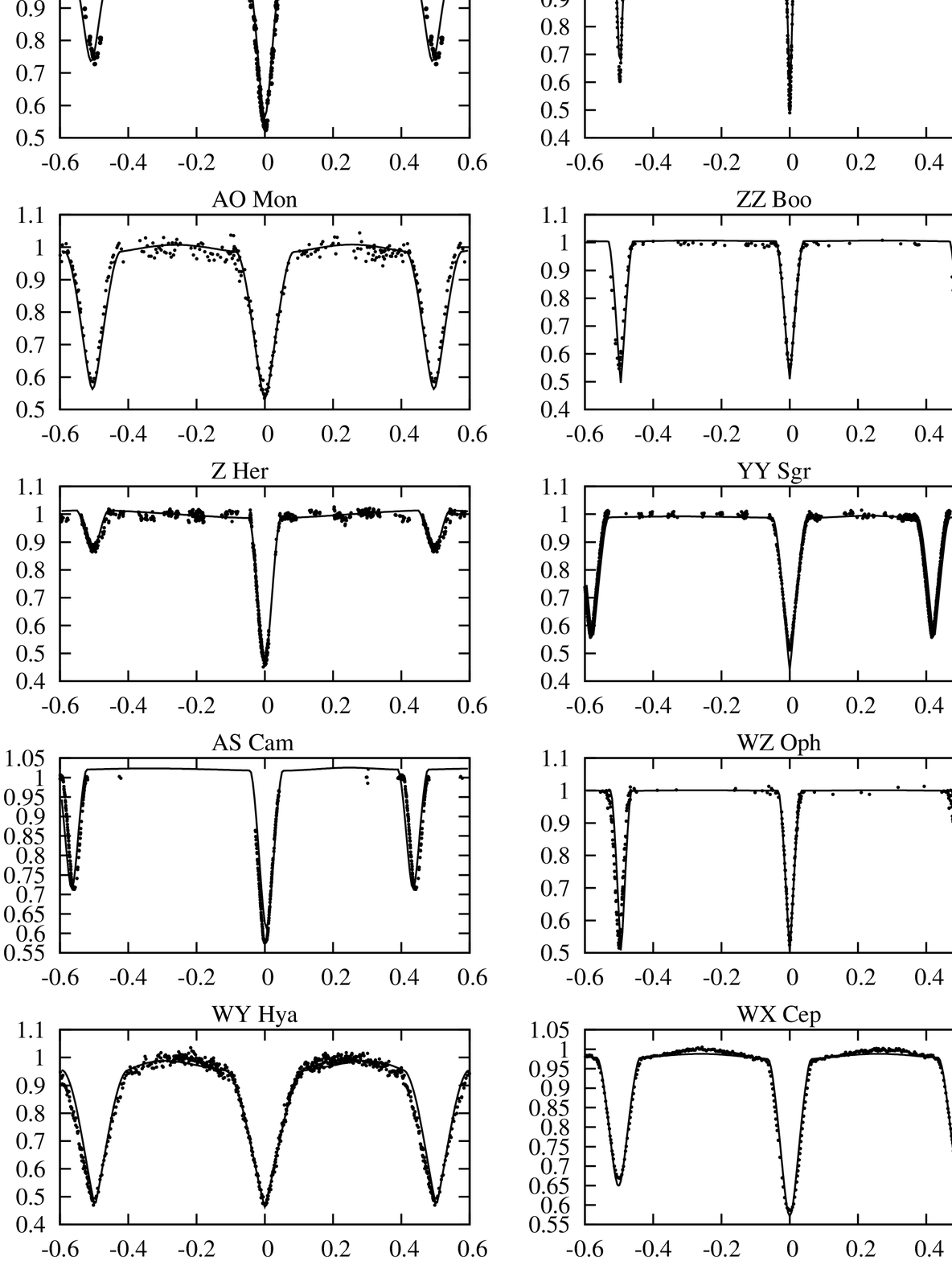} \\
\end{center}
\end{figure}

\begin{figure}
\caption{Continued: EBAI model light curves derived for the CALEB data.} \label{caleb_lcs4}
\begin{center}
\includegraphics[width=\textwidth]{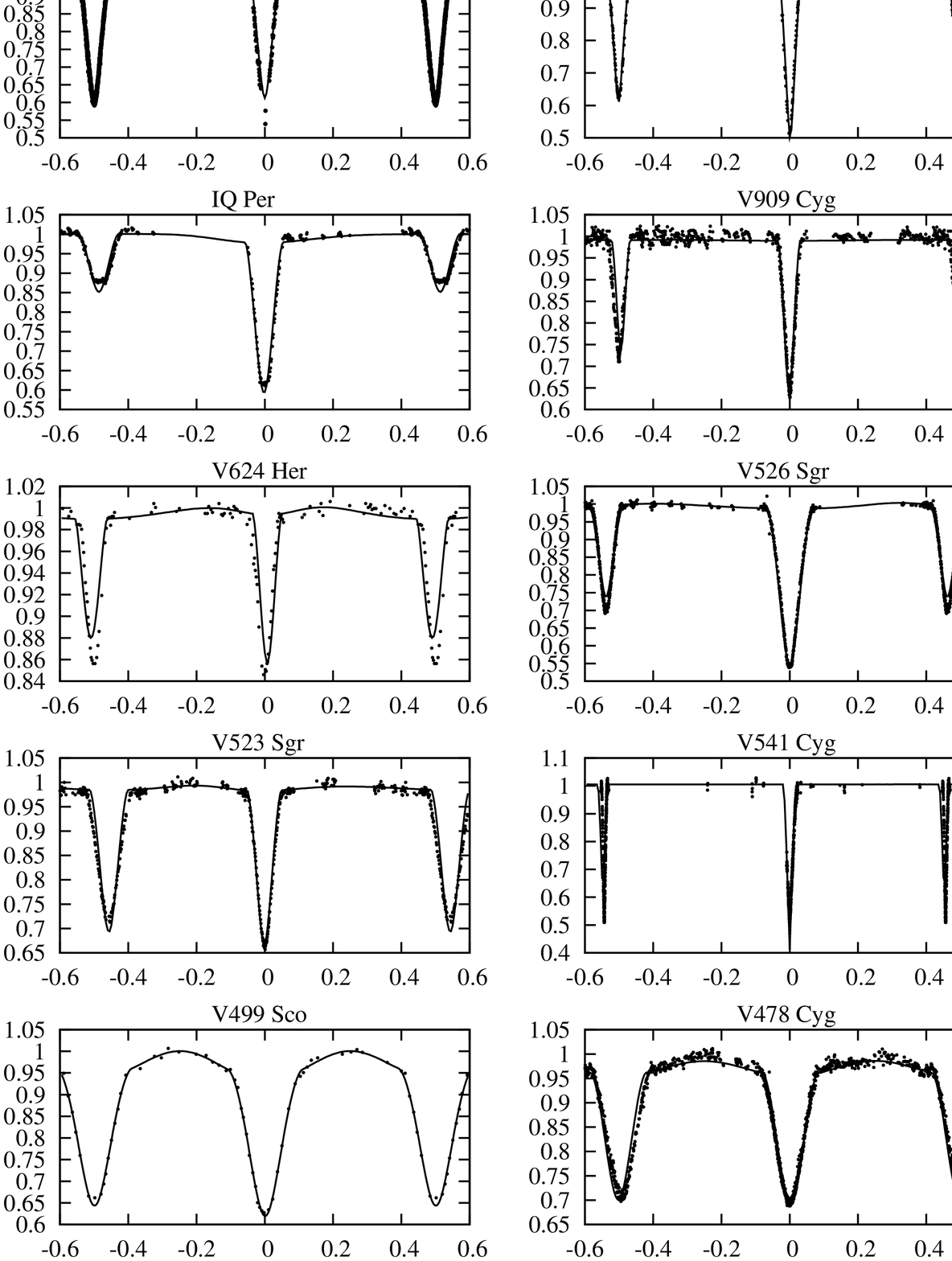} \\
\end{center}
\end{figure}

\begin{figure}
\caption{Continued: EBAI model light curves derived for the CALEB data.} \label{caleb_lcs5}
\begin{center}
\includegraphics[width=\textwidth]{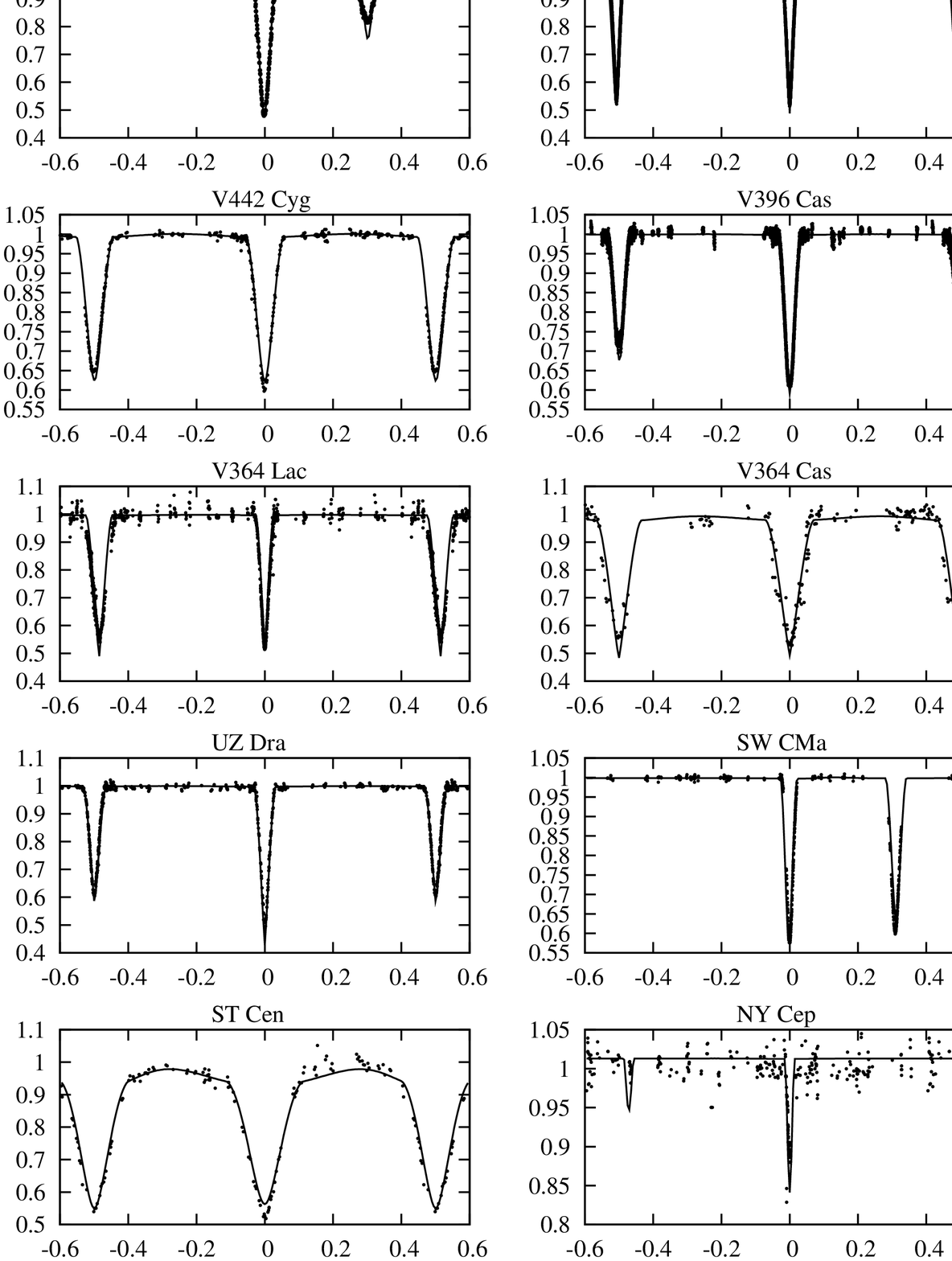} \\
\end{center}
\end{figure}


\begin{thebibliography}{}

\bibitem[\protect\astroncite{{Borucki} et~al.}{2004}]{borucki2004}
{Borucki}, W., {Koch}, D., {Boss}, A., {Dunham}, E., {Dupree}, A., {Geary}, J.,
  {Gilliland}, R., {Howell}, S., {Jenkins}, J., {Kondo}, Y., {Latham}, D.,
  {Lissauer}, J., and {Reitsema}, H.: 2004,
\newblock in F. {Favata}, S. {Aigrain}, and A. {Wilson} (eds.), {\em Stellar
  Structure and Habitable Planet Finding}, Vol. 538 of {\em ESA Special
  Publication}, pp 177--182

\bibitem[\protect\astroncite{{Devinney} et~al.}{2005}]{devinney2005}
{Devinney}, E., {Guinan}, E., {Bradstreet}, D., {DeGeorge}, M., {Giammarco},
  J., {Alcock}, C., and {Engle}, S.: 2005,
\newblock in {\em Bulletin of the American Astronomical Society}, Vol.~37, pp 1212--+

\bibitem[\protect\astroncite{{Devinney} et~al.}{2006}]{devinney2006}
{Devinney}, E.~J., {Guinan}, E., {DeGeorge}, M., {Bradstreet}, D.~H., and
  {Giammarco}, J.~M.: 2006,
\newblock in {\em Bulletin of the American Astronomical Society}, Vol.~38, pp 119--+

\bibitem[\protect\astroncite{{Devor}}{2005}]{devor2005}
{Devor}, J.: 2005,
\newblock {\em \apj} {\bf 628}, 411

\bibitem[\protect\astroncite{{Emery} and {Thomson}}{2001}]{emery2001}
{Emery}, W.~J. and {Thomson}, R.~E.: 2001,
\newblock {\em Data Analysis Methods in Physical Oceanography},
\newblock Elsevier

\bibitem[\protect\astroncite{{Faccioli} et~al.}{2008}]{faccioli2008}
{Faccioli}, L., {Alcock}, C., and {Cook}, K.: 2008,
\newblock {\em \apj} {\bf 672}, 604

\bibitem[\protect\astroncite{Freeman and Skapura}{1991}]{freeman1991}
Freeman, J.~A. and Skapura, D.~M.: 1991,
\newblock {\em Neural networks: algorithms, applications, and programming
  techniques},
\newblock Addison Wesley Longman Publishing Co., Inc., Redwood City, CA, USA

\bibitem[\protect\astroncite{{Guinan} et~al.}{2007}]{guinan2007}
{Guinan}, E.~F., {Engle}, S.~G., and {Devinney}, E.~J.: 2007,
\newblock in O. {Demircan}, S.~O. {Selam}, and B. {Albayrak} (eds.), {\em Solar
  and Stellar Physics Through Eclipses}, Vol. 370 of {\em Astronomical Society
  of the Pacific Conference Series}, pp 125--+

\bibitem[\protect\astroncite{{Guinan} et~al.}{2000}]{guinan2000}
{Guinan}, E.~F., {Ribas}, I., {Fitzpatrick}, E.~L., {Gim{\'e}nez}, {\'A}.,
  {Jordi}, C., {McCook}, G.~P., and {Popper}, D.~M.: 2000,
\newblock {\em \apj} {\bf 544}, 409

\bibitem[\protect\astroncite{{Kaiser} et~al.}{2002}]{kaiser2002}
{Kaiser}, N., {Aussel}, H., {Burke}, B.~E., {Boesgaard}, H., {Chambers}, K.,
  {Chun}, M.~R., {Heasley}, J.~N., {Hodapp}, K.-W., {Hunt}, B., {Jedicke}, R.,
  {Jewitt}, D., {Kudritzki}, R., {Luppino}, G.~A., {Maberry}, M., {Magnier},
  E., {Monet}, D.~G., {Onaka}, P.~M., {Pickles}, A.~J., {Rhoads}, P.~H.~H.,
  {Simon}, T., {Szalay}, A., {Szapudi}, I., {Tholen}, D.~J., {Tonry}, J.~L.,
  {Waterson}, M., and {Wick}, J.: 2002,
\newblock in J.~A. {Tyson} and S. {Wolff} (eds.), {\em Survey and Other
  Telescope Technologies and Discoveries. Edited by Tyson, J. Anthony; Wolff,
  Sidney. Proceedings of the SPIE, Volume 4836, pp. 154-164 (2002).}, pp
  154--164

\bibitem[\protect\astroncite{{Lucy}}{1967}]{lucy1967}
{Lucy}, L.~B.: 1967,
\newblock {\em Zeitschrift fur Astrophysics} {\bf 65}, 89

\bibitem[\protect\astroncite{{Mazeh} et~al.}{2006}]{mazeh2006}
{Mazeh}, T., {Tamuz}, O., and {North}, P.: 2006,
\newblock {\em \mnras} {\bf 367}, 1531

\bibitem[\protect\astroncite{{Michalska} and {Pigulski}}{2005}]{michalska2005}
{Michalska}, G. and {Pigulski}, A.: 2005,
\newblock {\em \aap} {\bf 434}, 89

\bibitem[\protect\astroncite{{Neelakanta} and {De
  Groff}}{1994}]{neelakanta1994}
{Neelakanta}, P.~S. and {De Groff}, D.~F.: 1994,
\newblock {\em Neural Network Modeling: Statistical Mechanics and Cybernetic
  Perspectives},
\newblock CRC Press

\bibitem[\protect\astroncite{{North} and {Zahn}}{2004}]{north2004}
{North}, P. and {Zahn}, J.-P.: 2004,
\newblock {\em New Astronomy Review} {\bf 48}, 741

\bibitem[\protect\astroncite{{Palanque-Delabrouille} et~al.}{1998}]{pd1998}
{Palanque-Delabrouille}, N., {Afonso}, C., {Albert}, J.~N., {Andersen}, J.,
  {Ansari}, R., {Aubourg}, E., {Bareyre}, P., {Bauer}, F., {Beaulieu}, J.~P.,
  {Bouquet}, A., {Char}, S., {Charlot}, X., {Couchot}, F., {Coutures}, C.,
  {Derue}, F., {Ferlet}, R., {Glicenstein}, J.~F., {Goldman}, B., {Gould}, A.,
  {Graff}, D., {Gros}, M., {Haissinski}, J., {Hamilton}, J.~C., {Hardin}, D.,
  {de Kat}, J., {Lesquoy}, E., {Loup}, C., {Magneville}, C., {Mansoux}, B.,
  {Marquette}, J.~B., {Maurice}, E., {Milsztajn}, A., {Moniez}, M.,
  {Perdereau}, O., {Prevot}, L., {Renault}, C., {Rich}, J., {Spiro}, M.,
  {Vidal-Madjar}, A., {Vigroux}, L., {Zylberajch}, S., and {The EROS
  Collaboration}: 1998,
\newblock {\em \aap} {\bf 332}, 1

\bibitem[\protect\astroncite{{Perryman} et~al.}{2001}]{perryman2001}
{Perryman}, M.~A.~C., {de Boer}, K.~S., {Gilmore}, G., {H{\o}g}, E.,
  {Lattanzi}, M.~G., {Lindegren}, L., {Luri}, X., {Mignard}, F., {Pace}, O.,
  and {de Zeeuw}, P.~T.: 2001,
\newblock {\em \aap} {\bf 369}, 339

\bibitem[\protect\astroncite{{Perryman} and {ESA}}{1997}]{perryman1997}
{Perryman}, M.~A.~C. and {ESA}: 1997,
\newblock {\em {The HIPPARCOS and TYCHO catalogues. Astrometric and photometric
  star catalogues derived from the ESA HIPPARCOS Space Astrometry Mission}},
\newblock The Hipparcos and Tycho catalogues.~Astrometric and photometric star
  catalogues derived from the ESA Hipparcos Space Astrometry Mission,
  Publisher: Noordwijk, Netherlands: ESA Publications Division, 1997, Series:
  ESA SP Series vol no: 1200, ISBN: 9290923997 (set)

\bibitem[\protect\astroncite{{Pojmanski}}{2002}]{pojmanski2002}
{Pojmanski}, G.: 2002,
\newblock {\em Acta Astronomica} {\bf 52}, 397

\bibitem[\protect\astroncite{{Popper} and {Etzel}}{1981}]{popper1981}
{Popper}, D.~M. and {Etzel}, P.~B.: 1981,
\newblock {\em \aj} {\bf 86}, 102

\bibitem[\protect\astroncite{{Pr{\v s}a} and {Zwitter}}{2005}]{phoebe}
{Pr{\v s}a}, A. and {Zwitter}, T.: 2005,
\newblock {\em \apj} {\bf 628}, 426

\bibitem[\protect\astroncite{{Pr\v sa} and {Zwitter}}{2005}]{prsa2005}
{Pr\v sa}, A. and {Zwitter}, T.: 2005,
\newblock in C. {Turon}, K.~S. {O'Flaherty}, and M.~A.~C. {Perryman} (eds.),
  {\em ESA SP-576: The Three-Dimensional Universe with Gaia}, pp 611--+

\bibitem[\protect\astroncite{{Pr{\v s}a} and {Zwitter}}{2007}]{prsa2006}
{Pr{\v s}a}, A. and {Zwitter}, T.: 2007,
\newblock in O. {Demircan}, S.~O. {Selam}, and B. {Albayrak} (eds.), {\em Solar
  and Stellar Physics Through Eclipses}, Vol. 370 of {\em Astronomical Society
  of the Pacific Conference Series}, pp 175--+

\bibitem[\protect\astroncite{{Pr{\v s}a} and {Zwitter}}{2006}]{prsa2007}
{Pr{\v s}a}, A. and {Zwitter}, T.: 2006,
\newblock in W. {Hartkopf}, P. {Harmanec}, and E.~F. {Guinan} (eds.), {\em IAU
  Symposium S240}, {Cambridge UP}

\bibitem[\protect\astroncite{{Tamuz} et~al.}{2006}]{tamuz2006}
{Tamuz}, O., {Mazeh}, T., and {North}, P.: 2006,
\newblock {\em \mnras} {\bf 367}, 1521

\bibitem[\protect\astroncite{{Tyson}}{2002}]{tyson2002}
{Tyson}, J.~A.: 2002,
\newblock in J.~A. {Tyson} and S. {Wolff} (eds.), {\em Survey and Other
  Telescope Technologies and Discoveries. Edited by Tyson, J. Anthony; Wolff,
  Sidney. Proceedings of the SPIE, Volume 4836, pp. 10-20 (2002).}, Vol. 4836
  of {\em Presented at the Society of Photo-Optical Instrumentation Engineers
  (SPIE) Conference}, pp 10--20

\bibitem[\protect\astroncite{{Udalski} et~al.}{1997}]{udalski1997}
{Udalski}, A., {Kubiak}, M., and {Szymanski}, M.: 1997,
\newblock {\em Acta Astronomica} {\bf 47}, 319

\bibitem[\protect\astroncite{{von Zeipel}}{1924}]{vonzeipel1924}
{von Zeipel}, H.: 1924,
\newblock {\bf 84}, 665

\bibitem[\protect\astroncite{{van Hamme}}{1993}]{vanhamme1993}
{van Hamme}, W.: 1993,
\newblock {\bf 106}, 2096

\bibitem[\protect\astroncite{{Wilson}}{1993}]{wd1993}
{Wilson}, R.~E.: 1993,
\newblock in K.-C. {Leung} and I.-S. {Nha} (eds.), {\em New Frontiers in Binary
  Star Research}, Vol.~38 of {\em Astronomical Society of the Pacific
  Conference Series}, pp 91--+

\bibitem[\protect\astroncite{{Wilson} and {Devinney}}{1971}]{wd1971}
{Wilson}, R.~E. and {Devinney}, E.~J.: 1971,
\newblock {\em \apj} {\bf 166}, 605

\bibitem[\protect\astroncite{{Wilson} and {Van Hamme}}{2007}]{wd2007}
{Wilson}, R.~E. and {Van Hamme}, W.: 2007,
\newblock {\em Computing Binary Star Observables},
\newblock University of Florida, Gainesville

\bibitem[\protect\astroncite{{Wyithe} and {Wilson}}{2001}]{ww2001}
{Wyithe}, J.~S.~B. and {Wilson}, R.~E.: 2001,
\newblock {\em \apj} {\bf 559}, 260

\bibitem[\protect\astroncite{{Wyithe} and {Wilson}}{2002}]{ww2002}
{Wyithe}, J.~S.~B. and {Wilson}, R.~E.: 2002,
\newblock {\em \apj} {\bf 571}, 293

\bibitem[\protect\astroncite{{Wyrzykowski} et~al.}{2003}]{wyrzykowski2003}
{Wyrzykowski}, L., {Udalski}, A., {Kubiak}, M., {Szymanski}, M., {Zebrun}, K.,
  {Soszynski}, I., {Wozniak}, P.~R., {Pietrzynski}, G., and {Szewczyk}, O.:
  2003,
\newblock {\em Acta Astronomica} {\bf 53}, 1

\end{thebibliography}
\end{document}